\documentclass[11pt,oneside]{article}%
\usepackage{cite}
\usepackage{amsmath}
\usepackage{amsfonts}
\usepackage{amssymb}
\usepackage{graphicx}%
\allowdisplaybreaks
\begin{document}

\author{Dzo Mikulovic\thanks{e-mail:dzo@theorie.physik.uni-muenchen.de}, Alexander
Schmidt\thanks{e-mail:schmidt@theorie.physik.uni-muenchen.de}, Hartmut
Wachter\thanks{e-mail:Hartmut.Wachter@physik.uni-muenchen.de}\\Sektion Physik, Ludwig-Maximilians-Universit\"{a}t,\\Theresienstr. 37, D-80333 M\"{u}nchen, Germany}
\title{Grassmann variables on quantum spaces}
\maketitle
\date{}

\begin{abstract}
Attention is focused on antisymmetrized \ versions of quantum spaces that are
of particular importance in physics, i.e. two-dimensional quantum plane,
q-deformed Euclidean space in three or four dimensions as well as q-deformed
Minkowski space. For each case standard techniques for dealing with q-deformed
Grassmann variables are developed. Formulae for multiplying supernumbers are
given. The actions of symmetry generators and fermionic derivatives upon
antisymmetrized quantum spaces are calculated. The complete Hopf structure for
all types of quantum space generators is written down. From the formulae for
the coproduct a realization of the $L$-matrices in terms of symmetry
generators can be read off. The $L$-matrices together with the action of
symmetry generators determine how quantum spaces of different type have to be
fused together.

\end{abstract}

\section{Introduction}

It is an old idea that limiting the precision of position measurements by a
fundamental length will lead to a new method for regularizing quantum field
theories \cite{Heis38}. It is also well-known that such a modification of
classical spacetime will in general break its Poincar\'{e} symmetry
\cite{Sny47}. One way out of this difficulty is to change not only spacetime,
but also its underlying symmetry.

Quantum groups can be seen as deformations of classical spacetime symmetries,
as they describe the symmetry of their comodules, the so-called quantum
spaces. From a physical point of view the most realistic examples for\ quantum
groups and quantum spaces arise from q-deformation \cite{Ku83, Dri85, Drin86,
Jim85, Wor87, Man88, RFT90}. In our work we are interested in q-deformed
versions of Minkowski space and Euclidean spaces as well as their
corresponding symmetries, given by q-deformed Lorentz algebra and algebras of
q-deformed angular momentum, respectively\ \cite{CSSW90, Pod90, SWZ91, Maj91,
LWW97}. Julius Wess and his coworkers were able to show that q-deformation of
spaces and symmetries can indeed lead to discretizations, as they result from
the existence of a smallest distance \cite{Fich97, CW98}. This observation
nourishes the hope that q-deformation might give a new method to regularize
quantum field theories \cite{MajReg, GKP96, Oec99, Blo03}.

In our previous work \cite{WW01, BW01, Wac02, Wac04,WacTr, WacBr} attention
was focused on symmetrized versions of q-deformed quantum spaces that are of
particular importance in physics, i.e. two-dimensional Manin plane, q-deformed
Euclidean space in three or four dimensions, and q-deformed Minkowski space.
As there is a need for Grassmann variables in physics we would like to discuss
antisymmetrized versions of those quantum spaces as well.

In particular, we intend to proceed as follows. In Sec. \ref{BasSec} we cover
the ideas our considerations about q-deformed quantum spaces are based on. For
further details we recommend Refs.\ \cite{Maj95} and \cite{ChDe96}. In the
subsequent sections we apply these reasonings to antisymmetrized versions of
two-dimensional quantum plane, q-deformed Euclidean space with three or four
dimensions as well as q-deformed Minkowski-space.

More concretely, we develop some standard techniques for dealing with
q-deformed Grassmann variables. In doing so, we start from the commutation
relations for q-deformed Grassmann variables and introduce q-deformed
supernumbers. After that we are going to derive explicit formulae for
multiplying q-deformed supernumbers. In addition to this, we are going to
calculate the action of symmetry generators and partial derivatives upon
antisymmetrized quantum spaces. Furthermore, we are going to write down the
complete Hopf structure on quantum space generators, including their
coproduct, antipode, and counit.

One should realize that the explicit form of the coproduct on quantum space
generators enables us to read off a realization of the so-called L-matrices in
terms of symmetry generators. This knowledge together with the action of
symmetry generators upon quantum spaces tells us how quantum spaces of
different type have to be fused together.

\section{ Basic ideas on antisymmetrized quantum spaces\label{BasSec}}

In our approach spacetime symmetries are described by quantum algebras like
$U_{q}(su_{2}),$ $U_{q}(so_{4})$ or q-deformed Lorentz algebra. Important for
us is the fact that these algebras are \textit{quasitriangular Hopf algebras},
i.e. their coproduct can be twisted by an invertible element $\mathcal{R}\in
H\otimes H$, which is known as the universal R-matrix of the corresponding
Hopf algebra $H$. Formally, we have%
\begin{equation}
\tau\circ\Delta h=\mathcal{R}(\Delta h)\mathcal{R}^{-1},\quad h\in H,
\end{equation}
where $\Delta$ and $\tau$ denote respectively the coproduct on $H$ and the
transposition map.

The modules of the quantum algebras are called \textit{quantum spaces}. At a
first glance a quantum space is nothing other than an algebra $A$ generated by
non-commuting coordinates $X_{1},X_{2},\ldots,X_{n},$ i.e.%

\begin{equation}
A=\mathbb{C}\left[  \left[  X_{1},\ldots X_{n}\right]  \right]  /\mathcal{I},
\end{equation}
where $\mathcal{I}$ denotes the ideal generated by the relations of the
non-commuting coordinates.

It should be noted that we can combine a quantum algebra $H$ with its
representation space $A$ to form a \textit{left cross product algebra}
$A\rtimes H$ built on $A\otimes H$ with product%
\begin{equation}
(a\otimes h)(b\otimes g)=a(h_{(1)}\triangleright b)\otimes h_{(2)}g,\quad
a,b\in A,\mathcal{\quad}h,g\in H,\label{LefCrosPro}%
\end{equation}
where $\triangleright$ denotes the left action of $H$ on $A.$ There is also a
right-handed version of this notion called a \textit{right cross product
algebra} $H\ltimes A$ and built on $H\otimes A$ with product%
\begin{equation}
(h\otimes a)(g\otimes b)=hg_{(2)}\otimes(a\triangleleft g_{(1)}%
)b,\label{RigCrosPro}%
\end{equation}
where $\triangleleft$ now stands for the right action of $H$ on $A.$ The last
two identities tell us that the commutation relations between symmetry
generators and representation space elements are completely determined by
coproduct and action of the symmetry generators, since we obtain from them%
\begin{equation}
hb=(h_{(1)}\triangleright b)h_{(2)},\quad ag=g_{(2)}(a\triangleleft g_{(1)}).
\end{equation}

However, in what follows it is necessary to take another point of view which
is provided by category theory. A category is a collection of objects
$X,Y,Z,\ldots$ together with a set Mor$(X,Y)$ of morphisms between two objects
$X,Y$. The composition of morphisms has similar properties as the composition
of maps. We are interested in tensor categories. These categories have a
product, denoted $\otimes$ and called the tensor product. It admits several
'natural' properties such as associativity and existence of a unit object. For
a more formal treatment we refer to Refs. \cite{Maj95}, \cite{Maj91Kat,
Maj94Kat} or \cite{MaL74}. If the action of a quasitriangular Hopf algebra $H$
on the tensor product of two quantum spaces $X$ and $Y$ is defined by
\begin{equation}
h\triangleright(v\otimes w)=(h_{(1)}\triangleright v)\otimes(h_{(2)}%
\triangleright w)\in X\otimes Y,\quad h\in H,
\end{equation}
where the coproduct is written in the so-called Sweedler notation, i.e.
$\Delta(h)=h_{(1)}\otimes h_{(2)},$ then the representations (quantum spaces)
of the given Hopf algebra (quantum algebra) are the objects of a tensor
category. In this tensor category exist a number of morphisms of particular
importance that are covariant with respect to the Hopf algebra action. First
of all, for any pair of objects $X,Y$ there is an isomorphism $\Psi
_{X,Y}:X\otimes Y\rightarrow Y\otimes X$ such that $(g\otimes f)\circ
\Psi_{X,Y}=\Psi_{X^{\prime},Y^{\prime}}\circ(f\otimes g)$ for arbitrary
morphisms $f\in$ Mor$(X,X^{\prime})$ and $g\in$ Mor$(Y,Y^{\prime})$. In
addition to this one requires the hexagon axiom to hold. The hexagon axiom is
the validity of the two conditions
\begin{equation}
\Psi_{X,Z}\circ\Psi_{Y,Z}=\Psi_{X\otimes Y,Z},\quad\Psi_{X,Z}\circ\Psi
_{X,Y}=\Psi_{X,Y\otimes Z}.
\end{equation}
A tensor category equipped with such mappings $\Psi_{X,Y}$ for each pair of
objects $X,Y$ is called a \textit{braided tensor category}. The mappings
$\Psi_{X,Y}$ as a whole are often referred to as the braiding of the tensor
category. Furthermore, for any quantum space algebra $X$ in this category
there are morphisms $\Delta:X\rightarrow X\otimes X,$ $S:X\rightarrow X,$ and
$\varepsilon:X\rightarrow\mathbb{C}$ forming a braided Hopf algebra, i.e.
$\Delta,$ $S,$ and $\epsilon$ obey the usual axioms of a Hopf algebra, but now
as morphisms in the braided category.

It is well-known that for a quasitriangular Hopf algebra $H$ the category of
$H$-modules is braided, with
\begin{equation}
\Psi_{X,Y}(v\otimes w)=(\mathcal{R}^{(2)}\triangleright w)\otimes
(\mathcal{R}^{(1)}\triangleright v),\text{\quad}v\in X,w\in Y,
\end{equation}
where $\mathcal{R}=\mathcal{R}^{(1)}\otimes\mathcal{R}^{(2)}.$ In terms of
quantum space generators the above identity becomes%
\begin{equation}
\Psi_{X,Y}(X^{i}\otimes Y^{j})=\hat{R}_{kl}^{ij}\,Y^{k}\otimes X^{l},
\end{equation}
where summation over repeated indices is to be understood. Notice that the
matrix $\hat{R}$ describes nothing other than a linear mapping between vector
spaces spanned by tensor products of quantum space generators. In the cases
under consideration this mapping can be restricted to invariant subspaces. As
a consequence, $\hat{R}$ admits a projector decomposition of the general form
\cite{KS97, LSW94}%
\begin{equation}
\hat{R}=\alpha_{S}P_{S}+\alpha_{A}P_{A}+\alpha_{T}P_{T},
\end{equation}
where $\alpha_{S},$ $\alpha_{A},$ and $\alpha_{T}$ denote the corresponding
eigenvalues. The projectors $P_{S}$ and $P_{A}$ are quantum analogs of a
symmetrizer and an antisymmetrizer, respectively, while $P_{T}$ projects onto
a one-dimensional subspace generated by the quantum length.

The relations of the \textit{quantum symmetric space} are determined by
\cite{Wess00}%
\begin{equation}
(P_{A})_{kl}^{ij}\,X^{k}X^{l}=0, \label{DefRel}%
\end{equation}
and likewise for the \textit{quantum antisymmetric space} (\textit{q-deformed
Grassmann algebra}),%
\begin{equation}
(P_{S})_{kl}^{ij}\,\theta^{k}\theta^{l}=0,\quad(P_{T})_{kl}^{ij}\,\theta
^{k}\theta^{l}=0. \label{DefAntiSpa}%
\end{equation}
Alternatively, the two identities defining quantum antisymmetric space can be
combined in the following way:%
\begin{align}
\theta^{i}\theta^{j}  &  =\left(  (P_{S})_{kl}^{ij}+(P_{A})_{kl}^{ij}%
+(P_{T})_{kl}^{ij}\right)  \theta^{k}\theta^{l}\label{RelAntiSymQuan}\\
&  =(P_{A})_{kl}^{ij}\,\theta^{k}\theta^{l}=\alpha_{A}^{-1}\hat{R}_{kl}%
^{ij}\,\theta^{k}\theta^{l}.\nonumber
\end{align}

Let us mention that the quantum spaces obtained this way satisfy the so-called
\textit{Poincar\'{e}-Birkhoff-Witt property}, i.e. the dimension of a subspace
of homogenous polynomials should be the same as for the corresponding
classical variables. This property is the deeper reason why normal ordered
monomials again constitute a basis of q-deformed Grassmann algebras.
Consequently, each q-deformed supernumber can be represented in the general
form
\begin{equation}
f(\underline{\theta})=f^{\prime}+\sum f_{\underline{K}}\,\theta^{\,\underline
{K}}, \label{SupNumAllg}%
\end{equation}
where the $f$'s\ are arbitrary complex numbers and the $\theta^{\underline{K}%
}$ stand for\ a monomials of a given normal ordering.

Next, we want to deal with the covariant differential calculus on quantum
spaces \cite{WZ91, CSW91, Song92}. Such a differential calculus can be
established by introducing an exterior derivative $d$ with the usual
properties of nilpotency and Leibniz rule:
\begin{align}
d^{2}  &  =0,\label{ExtDeriv}\\
d(fg)  &  =(df)g+(-1)^{\left\vert f\right\vert }f(dg),\nonumber
\end{align}
where
\begin{equation}
\left\vert f\right\vert =%
\begin{cases}
0, & \text{if }f\text{ bosonical,}\\
1, & \text{if }f\text{ fermionical.}%
\end{cases}
\end{equation}

In addition to this, we require that the differentials of the coordinates,
\begin{equation}
\xi^{i}\equiv dX^{i},\quad\eta^{i}\equiv d\theta^{i},
\end{equation}
are subject to the relations
\begin{align}
(P_{S})_{kl}^{ij}\,\xi^{k}\xi^{l}  &  =0,\quad(P_{T})_{kl}^{ij}\,\xi^{k}%
\xi^{l}=0,\\[0.1in]
(P_{A})_{kl}^{ij}\,\eta^{k}\eta^{l}  &  =0,\quad(P_{T})_{kl}^{ij}\,\eta
^{k}\eta^{l}=0.
\end{align}
With the same reasonings already applied in (\ref{RelAntiSymQuan}) the above
identities lead to
\begin{equation}
\xi^{i}\xi^{j}=\alpha_{A}^{-1}\hat{R}_{kl}^{ij}\,\xi^{k}\xi^{l},\quad\eta
^{i}\eta^{j}=\alpha_{S}^{-1}\hat{R}_{kl}^{ij}\,\eta^{k}\eta^{l}.
\label{DiffRelAlg}%
\end{equation}

In order to find commutation relations between coordinates and differentials,
we make as ansatz%
\begin{equation}
X^{i}\xi^{j}=B_{kl}^{ij}\,\xi^{k}X^{l},\text{\quad}\theta^{i}\eta^{j}%
=C_{kl}^{ij}\,\eta^{k}\theta^{l}. \label{VerDiffKoor}%
\end{equation}
Applying the exterior derivative to both sides of the above equations and
comparing the results with (\ref{DiffRelAlg}) then yields for the unknown
coefficients%
\begin{equation}
B_{kl}^{ij}=-\alpha_{A}^{-1}\hat{R}_{kl}^{ij},\quad C_{kl}^{ij}=\alpha
_{S}^{-1}\hat{R}_{kl}^{ij}. \label{BCKoef}%
\end{equation}

As a next step we introduce partial derivatives by
\begin{equation}
d=\xi^{i}(\partial_{x})_{i}\quad\text{and\quad}d=\eta^{i}(\partial_{\theta
})_{i}.
\end{equation}
From (\ref{ExtDeriv}) together with (\ref{VerDiffKoor}) it can be shown that
the following Leibniz rules hold:%
\begin{align}
(\partial_{x})_{i}X^{j}  &  =\delta_{i}^{j}-\alpha_{A}^{-1}\hat{R}_{il}%
^{jk}\,X^{l}(\partial_{x})_{k},\label{LeibRuleAllg}\\
(\partial_{\theta})_{i}\theta^{j}  &  =\delta_{i}^{j}-\alpha_{S}^{-1}\hat
{R}_{il}^{jk}\,\theta^{l}(\partial_{\theta})_{k}.\nonumber
\end{align}

We could also have started our considerations from the inverse braiding
\begin{equation}
\Psi_{X,Y}^{-1}(v\otimes w)=((\mathcal{R}^{-1})^{(1)}\triangleright
w)\otimes((\mathcal{R}^{-1})^{(2)}\triangleright v),
\end{equation}
leading us to%
\begin{equation}
\Psi_{X,Y}^{-1}(X^{i}\otimes Y^{j})=(\hat{R}^{-1})_{kl}^{ij}\,Y^{k}\otimes
X^{l}.
\end{equation}
$\hat{R}^{-1}$ denotes the inverse of $\hat{R}$, so its projector
decomposition is given by
\begin{equation}
\hat{R}^{-1}=\alpha_{S}^{-1}P_{S}+\alpha_{A}^{-1}P_{A}+\alpha_{T}^{-1}P_{T}.
\end{equation}
Repeating the same steps as before we get relations for conjugated objects.
However, their explicit form can be obtained from the above relations most
easily by applying the substitutions%
\begin{gather}
\hat{R}\rightarrow\hat{R}^{-1},\quad\alpha_{A,S,T}\rightarrow\alpha
_{A,S,T}^{-1},\\
\partial_{x}^{i}\rightarrow\hat{\partial}_{x}^{i},\quad\partial_{\theta}%
^{i}\rightarrow\hat{\partial}_{\theta}^{i},\nonumber\\
a^{i}\rightarrow\bar{a}^{i},\quad a^{i}\in\{\xi^{i},\eta^{i},X^{i},\theta
^{i}\}.\nonumber
\end{gather}

Lastly, let us say a few words about the Hopf structures on quantum spaces.
With the L-matrix and its conjugate \cite{Tan3}, which can be introduced by%
\begin{align}
\Psi_{X,Y}(a^{i}\otimes w)  &  =((\mathcal{L}_{a})_{j}^{i}\triangleright
w)\otimes a^{j},\\
\Psi_{X,Y}^{-1}(a^{i}\otimes w)  &  =((\mathcal{\bar{L}}_{a})_{j}%
^{i}\triangleright w)\otimes a^{j},\nonumber
\end{align}
the two Hopf structures on quantum space generators can be written as
\cite{Maj-Pr}%
\begin{align}
\Delta(a^{i})  &  =a^{i}\otimes1+(\mathcal{L}_{a})_{j}^{i}\otimes
a^{j},\label{HopfStruc}\\
\bar{\Delta}(a^{i})  &  =a^{i}\otimes1+(\mathcal{\bar{L}}_{a})_{j}^{i}\otimes
a^{j},\nonumber\\
S(a^{i})  &  =-S(\mathcal{L}_{a})_{j}^{i}\,a^{j},\label{Antipode}\\
\bar{S}(a^{i})  &  =-S(\mathcal{\bar{L}}_{a})_{j}^{i}\,a^{j},\nonumber\\
\varepsilon(a^{i})  &  =\bar{\varepsilon}(a^{i})=0.
\end{align}
One should notice that the entries of the L-matrices live in the corresponding
quantum algebra $H$ \ This way, we can conclude that the above expressions are
part of the Hopf structure of the crossed product algebra $A\rtimes H$.

\section{Two-dimensional quantum plane}

We begin by describing the two-dimensional antisymmetrized quantum plane
algebra explicitly. For this purpose we need the projector decomposition of
the R-matrix for $U_{q}(su_{2})$ \cite{KS97}:
\begin{equation}
\hat{R}=qP_{S}-q^{-1}P_{A}. \label{ProZerR}%
\end{equation}
One should notice that in this case $P_{A}$ and $P_{T}$ coincide, so we have
only two different projectors in Eq. (\ref{ProZerR}). For the antisymmetrized
coordinates, the decomposition in (\ref{ProZerR}) implies [cf. Eq.
(\ref{RelAntiSymQuan})]%
\begin{equation}
\theta^{i}\theta^{j}=-q\hat{R}_{kl}^{ij}\,\theta^{k}\theta^{l}.
\label{AnSymRel2dim}%
\end{equation}
Inserting the explicit form for the R-matrix \cite{LSW94}, we get from Eq.
(\ref{AnSymRel2dim})\ the following independent relations:%
\begin{equation}
(\theta^{1})^{2}=(\theta^{2})^{2}=0,\quad\theta^{1}\theta^{2}=-q^{-1}%
\theta^{2}\theta^{1}. \label{Ver2dimFer}%
\end{equation}

To go further, we introduce supernumbers, which we can write in the form
\begin{equation}
f(\theta^{1},\theta^{2})=f^{\prime}+f_{1}\theta^{1}+f_{2}\theta^{2}%
+f_{12}\theta^{1}\theta^{2}. \label{Supnum2dim}%
\end{equation}
Using relations (\ref{Ver2dimFer}) it is not very difficult to show that the
product of two supernumbers can be written as
\begin{align}
&  (f\cdot g)(\theta^{1},\theta^{2})\\
&  =(f\cdot g)^{\prime}+(f\cdot g)_{1}\,\theta^{1}+(f\cdot g)_{2}\,\theta
^{2}+(f\cdot g)_{12}\,\theta^{1}\theta^{2},\nonumber
\end{align}
with
\begin{align}
(f\cdot g)^{\prime}  &  =f^{\prime}g^{\prime},\\
(f\cdot g)_{i}  &  =f_{i}g^{\prime}+f^{\prime}g_{i},\quad i=1,2,\nonumber\\
(f\cdot g)_{12}  &  =f_{1}g_{2}-qf_{2}g_{1}+f^{\prime}g_{12}+f_{12}g{\prime
}.\nonumber
\end{align}

Next, we come to the action of symmetry generators on supernumbers. To this
end, we have to recall that both bosonic and fermionic coordinates transform
as spinors under the action of the symmetry algebra $U_{q}(su_{2}).$ Using for
$U_{q}(su_{2})$ the form as it was introduced in Ref. \cite{LWW97} the
commutation relations between its independent generators (denoted by $T^{+},$
$T^{-},$ and $\tau$) and the spinor components $a^{i}$, $i=1,2,$ read as
\begin{align}
T^{+}a^{1}  &  =qa^{1}T^{+}+q^{-1}a^{2},\label{RelTKoKanf}\\
T^{+}a^{2}  &  =q^{-1}a^{2}T^{+},\nonumber\\[0.16in]
T^{-}a^{1}  &  =qa^{1}T^{-},\\
T^{-}a^{2}  &  =q^{-1}a^{2}T^{-}+qa^{1},\nonumber\\[0.16in]
\tau a^{1}  &  =q^{2}a^{1}\tau,\label{RelTKoKend}\\
\tau a^{2}  &  =q^{-2}a^{2}\tau.\nonumber
\end{align}

From the above relations we can derive the action of the symmetry generators
on a supernumber of the form (\ref{Supnum2dim}). To this end, we repeatedly
apply the commutation relations (\ref{RelTKoKanf}) - (\ref{RelTKoKend}) to the
product of a symmetry generator and a supernumber, until we obtain an
expression with all symmetry generators standing to the right of all quantum
plane coordinates. In doing so, we get the left action of a symmetry generator
on a supernumber. Explicitly, we find
\begin{align}
T^{+}\triangleright f(\theta^{1},\theta^{2})  &  =q^{-1}f_{1}\theta^{2},\\
T^{-}\triangleright f(\theta^{1},\theta^{2})  &  =qf_{2}\theta^{1},\nonumber\\
\tau\triangleright f(\theta^{1},\theta^{2})  &  =f(q^{2}\theta^{1}%
,q^{-2}\theta^{2}).\nonumber
\end{align}

Right actions of symmetry generators on supernumbers can be derived in a
similar way, if we now consider a generator standing to the right of a
supernumber and commute it to the left of all quantum space coordinates.
Proceeding in this manner one can verify a remarkable correspondence between
right and left actions. More concretely, we have the transformations%
\begin{align}
f(\theta^{1},\theta^{2})\triangleleft T^{\pm}  &  \overset{i\leftrightarrow
i^{\prime}}{\longleftrightarrow}-q^{\mp3}T^{\mp}\triangleright f(\theta
^{1},\theta^{2}),\\
f(\theta^{1},\theta^{2})\triangleleft\tau &  \overset{i\leftrightarrow
i^{\prime}}{\longleftrightarrow}\tau\triangleright f(\theta^{1},\theta
^{2}),\nonumber
\end{align}
where the symbol $\overset{i\leftrightarrow i^{\prime}}{\longleftrightarrow}$
indicates the following transitions:
\begin{gather}
\theta^{i}\overset{i\leftrightarrow i^{\prime}}{\longleftrightarrow}%
\theta^{i^{\prime}},\quad\theta^{i}\theta^{j}\overset{i\leftrightarrow
i^{\prime}}{\longleftrightarrow}\theta^{j^{\prime}}\theta^{i^{\prime}},\\
f^{\prime}\overset{i\leftrightarrow i^{\prime}}{\longleftrightarrow}f^{\prime
},\quad f_{i}\overset{i\leftrightarrow i^{\prime}}{\longleftrightarrow
}f_{i^{\prime}},\quad f_{ij}\overset{i\leftrightarrow i^{\prime}%
}{\longleftrightarrow}f_{j^{\prime}i^{\prime}},\nonumber\\
i^{\prime}\equiv3-i,\quad i,j=1,2.\nonumber
\end{gather}
For this to become more clear, we give as an example
\begin{equation}
-q^{3}T^{+}\triangleright f(\theta^{1},\theta^{2})=-q^{2}f_{1}\theta
^{2}\overset{i\leftrightarrow i^{\prime}}{\longleftrightarrow}-q^{2}%
f_{2}\theta^{1}=f(\theta^{1},\theta^{2})\triangleleft T^{-}.
\end{equation}

Now, we turn our attention to the covariant differential calculus on the
quantum plane. The differentials of bosonical and fermionical coordinates are
subject to the relations [cf. Eq. (\ref{DiffRelAlg})]%
\begin{equation}
\xi^{i}\xi^{j}=-q\hat{R}_{kl}^{ij}\,\xi^{k}\xi^{l},\quad\eta^{i}\eta
^{j}=q^{-1}\hat{R}_{kl}^{ij}\,\eta^{k}\eta^{l}, \label{VerDiff}%
\end{equation}
which is consistent with [cf. Eqs. (\ref{VerDiffKoor}) and (\ref{BCKoef})]%
\begin{equation}
X^{i}\xi^{j}=q\hat{R}_{kl}^{ij}\,\xi^{k}X^{l},\quad\theta^{i}\eta^{j}%
=q^{-1}\hat{R}_{kl}^{ij}\,\eta^{k}\theta^{l}. \label{VerKoorDiff}%
\end{equation}

In what follows we need the q-deformed spinor metric $\varepsilon^{ij}$ and
its inverse $\varepsilon_{ij}$\ given by \cite{SS90}
\begin{equation}
\varepsilon^{11}=\varepsilon^{22}=0,\quad\varepsilon^{12}=-q^{-1/2}%
,\quad\varepsilon^{21}=q^{1/2},
\end{equation}
and
\begin{equation}
\varepsilon_{ij}=-\varepsilon^{ij}.
\end{equation}
Now, we are able to raise and lower spinor indices as usual, i.e.
\begin{equation}
a_{i}=\varepsilon_{ij}a^{j},\text{\quad}a^{i}=\varepsilon^{ij}a_{j}.
\end{equation}
With the identity%
\begin{equation}
\varepsilon^{nk}\,\hat{R}_{kl}^{ij}=q(\hat{R}^{-1})_{lk}^{ni}\,\varepsilon
^{kj},
\end{equation}
one may check that for partial derivatives with upper indices the Leibniz
rules in (\ref{LeibRuleAllg}) become%
\begin{align}
\partial_{x}^{i}X^{j}  &  =\varepsilon^{ij}+q^{2}(\hat{R}^{-1})_{kl}%
^{ij}\,X^{k}\partial_{x}^{l},\label{VerAK}\\
\partial_{\theta}^{i}\theta^{j}  &  =\varepsilon^{ij}-(\hat{R}^{-1})_{kl}%
^{ij}\,\theta^{k}\partial_{\theta}^{l}.\nonumber
\end{align}

Applying the substitutions
\begin{gather}
\partial_{a}\rightarrow\hat{\partial}_{a},\quad a\rightarrow\bar{a},\quad
a\in\{\xi,\eta,X,\theta\},\label{SubReg2dim}\\
q\rightarrow q^{-1},\quad\hat{R}\rightarrow\hat{R}^{-1},\nonumber
\end{gather}
to relations (\ref{VerDiff}), (\ref{VerKoorDiff}), and (\ref{VerAK}) then
yields the corresponding identities for the conjugated differential calculus.

Next, we want to deal on with the actions of partial derivatives on
supernumbers. This way, we can proceed in very much the same way as was done
in the case of symmetry generators. Written out explicitly, the relations in
(\ref{VerAK}) become for the fermionic case
\begin{align}
\partial_{\theta}^{1}\theta^{1}  &  =-q^{-1}\theta^{1}\partial_{\theta}%
^{1},\label{VerFerAbXAnf}\\
\partial_{\theta}^{1}\theta^{2}  &  =-q^{-1/2}-\theta^{2}\partial_{\theta}%
^{1},\nonumber\\[0.16in]
\partial_{\theta}^{2}\theta^{1}  &  =q^{1/2}-\theta^{1}\partial_{\theta}%
^{2}+\lambda\theta^{2}\partial_{\theta}^{1},\label{VerFer2dimAbxEnd}\\
\partial_{\theta}^{2}\theta^{2}  &  =-q^{-1}\theta^{2}\partial_{\theta}%
^{2},\nonumber
\end{align}
and likewise for the conjugated partial derivatives,%
\begin{align}
\hat{\partial}_{\theta}^{1}\bar{\theta}^{1}  &  =-q\bar{\theta}^{1}%
\hat{\partial}_{\theta}^{1},\\
\hat{\partial}_{\theta}^{1}\bar{\theta}^{1}  &  =-q^{-1/2}-\bar{\theta}%
^{2}\hat{\partial}_{\theta}^{1}-\lambda\bar{\theta}^{1}\hat{\partial}_{\theta
}^{2},\nonumber\\[0.16in]
\hat{\partial}_{\theta}^{1}\bar{\theta}^{1}  &  =q^{1/2}-\bar{\theta}^{1}%
\hat{\partial}_{\theta}^{2},\\
\hat{\partial}_{\theta}^{1}\bar{\theta}^{1}  &  =-q\bar{\theta}^{2}%
\hat{\partial}_{\theta}^{2},\nonumber
\end{align}
where $\lambda=q-q^{-1}$. From (\ref{VerFerAbXAnf}) and
(\ref{VerFer2dimAbxEnd}) it follows that
\begin{align}
\partial_{\theta}^{1}\triangleright f(\theta^{2},\theta^{1})  &
=-q^{-1/2}f_{2}-q^{-1/2}f_{21}\theta^{1},\\
\partial_{\theta}^{2}\triangleright f(\theta^{2},\theta^{1})  &  =q^{1/2}%
f_{1}-q^{-1/2}f_{21}\theta^{2}.\nonumber
\end{align}

Repeating the same steps as before for conjugated partial derivatives as well
as right actions one can verify the correspondences%
\begin{align}
\hat{\partial}_{\theta}^{i}\,\bar{\triangleright}\,f(\bar{\theta}^{1}%
,\bar{\theta}^{2})  &  \overset{{%
\genfrac{}{}{0pt}{}{i}{q}%
}{%
\genfrac{}{}{0pt}{}{\rightarrow}{\rightarrow}%
}{%
\genfrac{}{}{0pt}{}{i^{\prime}}{1/q}%
}}{\longleftrightarrow}-\partial_{\theta}^{i^{\prime}}\triangleright
f(\theta^{2},\theta^{1}),\\
q^{-1}f(\bar{\theta}^{2},\bar{\theta}^{1})\,\bar{\triangleleft}\,{\hat
{\partial}}_{\theta}^{i}  &  \overset{{%
\genfrac{}{}{0pt}{}{i}{q}%
}{%
\genfrac{}{}{0pt}{}{\rightarrow}{\rightarrow}%
}{%
\genfrac{}{}{0pt}{}{i^{\prime}}{1/q}%
}}{\longleftrightarrow}-qf(\theta^{1},\theta^{2})\triangleleft\partial
_{\theta}^{i^{\prime}},\nonumber
\end{align}
and
\begin{align}
f(\theta^{1},\theta^{2})\triangleleft\partial_{\theta}^{i}  &  \overset
{i\leftrightarrow i^{\prime}}{\longleftrightarrow}-q^{-1}\hat{\partial
}_{\theta}^{i^{\prime}}\,\bar{\triangleright}\,f(\bar{\theta}^{1},\bar{\theta
}^{2}),\\[0.16in]
f(\bar{\theta}^{2},\bar{\theta}^{1})\,\bar{\triangleleft}\,\hat{\partial
}_{\theta}^{i}  &  \overset{i\leftrightarrow i^{\prime}}{\longleftrightarrow
}-q\partial_{\theta}^{i^{\prime}}\triangleright f(\theta^{2},\theta
^{1}),\nonumber
\end{align}
where the symbol $\overset{{%
\genfrac{}{}{0pt}{}{i}{q}%
}{%
\genfrac{}{}{0pt}{}{\rightarrow}{\rightarrow}%
}{%
\genfrac{}{}{0pt}{}{i^{\prime}}{1/q}%
}}{\longleftrightarrow}$ now describes substitutions given by
\begin{gather}
\theta^{i}\overset{{%
\genfrac{}{}{0pt}{}{i}{q}%
}{%
\genfrac{}{}{0pt}{}{\rightarrow}{\rightarrow}%
}{%
\genfrac{}{}{0pt}{}{i^{\prime}}{1/q}%
}}{\longleftrightarrow}\theta^{i^{\prime}},\quad\theta^{i}\theta^{j}\overset{{%
\genfrac{}{}{0pt}{}{i}{q}%
}{%
\genfrac{}{}{0pt}{}{\rightarrow}{\rightarrow}%
}{%
\genfrac{}{}{0pt}{}{i^{\prime}}{1/q}%
}}{\longleftrightarrow}\theta^{i^{\prime}}\theta^{j^{\prime}},\\
f^{\prime}\overset{{%
\genfrac{}{}{0pt}{}{i}{q}%
}{%
\genfrac{}{}{0pt}{}{\rightarrow}{\rightarrow}%
}{%
\genfrac{}{}{0pt}{}{i^{\prime}}{1/q}%
}}{\longleftrightarrow}f^{\prime},\quad f_{i}\overset{{%
\genfrac{}{}{0pt}{}{i}{q}%
}{%
\genfrac{}{}{0pt}{}{\rightarrow}{\rightarrow}%
}{%
\genfrac{}{}{0pt}{}{i^{\prime}}{1/q}%
}}{\longleftrightarrow}f_{i^{\prime}},\quad f_{ij}\overset{{%
\genfrac{}{}{0pt}{}{i}{q}%
}{%
\genfrac{}{}{0pt}{}{\rightarrow}{\rightarrow}%
}{%
\genfrac{}{}{0pt}{}{i^{\prime}}{1/q}%
}}{\longleftrightarrow}f_{i^{\prime}j^{\prime}},\nonumber\\
q\overset{{%
\genfrac{}{}{0pt}{}{i}{q}%
}{%
\genfrac{}{}{0pt}{}{\rightarrow}{\rightarrow}%
}{%
\genfrac{}{}{0pt}{}{i^{\prime}}{1/q}%
}}{\longleftrightarrow}q^{-1},\quad i,j=1,2.\nonumber
\end{gather}
It should be noticed that the normal ordering the\ representation of a
supernumber refers to is indicated by the order in which arguments are
arranged in the symbol for the supernumber (see also Appendix \ref{RepSup}).

Now, we come to the Hopf structure on quantum space generators. A short glance
at (\ref{HopfStruc}) and (\ref{Antipode}) shows us that the explicit form of
coproduct and antipode is completely determined by the L-matrices. Therefore,
our task is to find for the unknown entries of the L-matrix combinations of
symmetry generators that produce the correct commutation relations between
generators of different quantum spaces. In other words, exploiting the
identities%
\begin{equation}
a^{i}b^{j}=\left(  (\mathcal{L}_{a})_{k}^{i}\triangleright b^{j}\right)
\,a^{k},\quad b^{i}a^{j}=\left(  (\mathcal{\bar{L}}_{b})_{k}^{i}\triangleright
a^{j}\right)  \,b^{k}, \label{BraidRel}%
\end{equation}
we should be able to regain the relations in (\ref{VerDiff}),
(\ref{VerKoorDiff}), (\ref{VerAK}), and their conjugated counterparts (if
inhomogeneous terms are discounted). We have found L-matrices with this
property. Inserting their explicit form into Eqs. (\ref{HopfStruc}) and
(\ref{Antipode}) we get for coproduct, antipode, and counit the expressions%
\begin{align}
\Delta(a^{1})  &  =a^{1}\otimes1+\Lambda(a)\tau^{-1/4}\otimes a^{1}%
,\label{HopfZwei}\\
\Delta(a^{2})  &  =a^{2}\otimes1+\Lambda(a)\tau^{1/4}\otimes a^{2}%
-q\lambda\Lambda(a)\tau^{-1/4}T^{+}\otimes a^{1},\nonumber\\[0.16in]
S(a^{1})  &  =-\Lambda^{-1}(a)\tau^{1/4}a^{1},\\
S(a^{2})  &  =-\Lambda^{-1}(a)\tau^{-1/4}a^{2}-q^{2}\lambda\Lambda^{-1}%
(a)\tau^{-1/4}T^{+}a^{1},\nonumber\\[0.16in]
\varepsilon(a^{1})  &  =\varepsilon(a^{2})=0,
\end{align}
and similarly for the Hopf structure to the conjugated L-matrix,%
\begin{align}
\bar{\Delta}(a^{1})  &  =a^{1}\otimes1+\Lambda^{-1}(a)\tau^{1/4}\otimes
a^{1}+q^{-1}\lambda\Lambda^{-1}(a)\tau^{-1/4}T^{-}\otimes a^{2}%
,\label{HopfZweiKon}\\
\bar{\Delta}(a^{2})  &  =a^{2}\otimes1+\Lambda^{-1}(a)\tau^{-1/4}\otimes
a^{2},\nonumber\\[0.16in]
\bar{S}(a^{1})  &  =-\Lambda(a)\tau^{-1/4}a^{1}+q^{-2}\lambda\Lambda
(a)\tau^{-1/4}T^{-}a^{2},\\
\bar{S}(a^{2})  &  =-\Lambda(a)\tau^{1/4}a^{2},\nonumber\\[0.16in]
\bar{\varepsilon}(a^{1})  &  =\bar{\varepsilon}(a^{2})=0,
\end{align}
where $a$ stands for one of the following quantities
\begin{equation}
a\in\{\partial_{x},\partial_{\theta},X,\theta,\xi,\eta\}.
\end{equation}

From (\ref{HopfZwei}) and (\ref{HopfZweiKon}) we can see that the L-matrices
depend on unitary scaling operators denoted by $\Lambda(a).$ To understand the
occurrence of these scaling operators we have to take a look at the
commutation relations in (\ref{VerKoorDiff}) and (\ref{VerAK}), which tell us
that the braiding between generators of different quantum spaces is given by
the R-matrix or its inverse up to a constant factor. The point now is that the
action of the scaling operators have been determined in such a way that the
relations in (\ref{BraidRel})\ lead to the correct factors if we consider the
braiding between generators of different quantum spaces. This can be achieved
by specifying the scaling operators according to
\begin{equation}
\Lambda(\partial_{x}^{i})=\Lambda^{-3/4},\quad\Lambda(X^{i})=\Lambda
^{3/4},\quad\Lambda(\eta^{i})=\Lambda^{-1/4} \label{lambda1}%
\end{equation}
and
\begin{equation}
\Lambda(\partial_{\theta}^{i})=\tilde{\Lambda}^{-3},\quad\Lambda(\theta
^{i})=\tilde{\Lambda}^{3},\quad\Lambda(\xi^{i})=\tilde{\Lambda}^{3},
\label{lambdatilde1}%
\end{equation}
where the grouplike operators $\Lambda$ and $\tilde{\Lambda}$ satisfy the
commutation relations%
\begin{align}
\Lambda X^{i}  &  =q^{-2}X^{i}\Lambda, & \tilde{\Lambda}X^{i}  &
=q^{3/2}X^{i}\tilde{\Lambda},\\
\Lambda\partial_{x}^{i}  &  =q^{2}\partial_{x}^{i}\Lambda, & \tilde{\Lambda
}\partial_{x}^{i}  &  =q^{-3/2}\partial_{x}^{i}\tilde{\Lambda},\nonumber\\
\Lambda\xi^{i}  &  =q^{-2}\xi^{i}\Lambda, & \tilde{\Lambda}\xi^{i}  &
=-q^{-1/2}\xi^{i}\tilde{\Lambda},\nonumber\\
\Lambda\eta^{i}  &  =q^{\frac{2}{3}}\eta^{i}\Lambda, & \tilde{\Lambda}%
\eta^{i}  &  =q^{-1/2}\eta^{i}\tilde{\Lambda},\nonumber\\
\Lambda\theta^{i}  &  =q^{-2}\theta^{i}\Lambda, & \tilde{\Lambda}\theta^{i}
&  =-q^{-1/2}\theta^{i}\tilde{\Lambda},\nonumber\\
\Lambda\partial_{\theta}^{i}  &  =q^{2}\partial_{\theta}^{i}\Lambda, &
\tilde{\Lambda}\partial_{\theta}^{i}  &  =-q^{1/2}\partial_{\theta}^{i}%
\tilde{\Lambda}.\nonumber
\end{align}
Finally, let us notice that the above identities for the scaling operator have
been derived by exploiting consistence arguments like%
\begin{align}
a^{i}b^{j}  &  =\left(  (\mathcal{L}_{a})_{k}^{i}\triangleright b^{j}\right)
\,a^{k}=b^{k}\,(a^{i}\triangleleft(\mathcal{\bar{L}}_{b})_{k}^{j}),\\
b^{i}a^{j}  &  =\left(  (\mathcal{\bar{L}}_{b})_{k}^{i}\triangleright
a^{j}\right)  \,b^{k}=a^{k}\,(b^{i}\triangleleft(\mathcal{L}_{a})_{k}%
^{j}).\nonumber
\end{align}

\section{Three-dimensional Euclidean space}

All considerations of the previous sections pertain equally to the
three-dimensional q-deformed Euclidean space \cite{LWW97}. Thus we can
restrict ourselves to stating the results, only. Now, the projector
decomposition of the R-matrix becomes
\begin{equation}
\hat{R}=P_{S}-q^{-4}P_{A}+q^{-6}P_{T}.
\end{equation}
The relations for the fermionic\ quantum space coordinates are given by
\begin{equation}
\theta^{A}\theta^{B}=-q^{4}\hat{R}_{CD}^{AB}\,\theta^{C}\theta^{D},
\end{equation}
which is equivalent to the following independent relations:
\begin{align}
(\theta^{+})^{2}  &  =(\theta^{-})^{2}=0,\\
\theta^{3}\theta^{3}  &  =\lambda\,\theta^{+}\theta^{-},\nonumber\\
\theta^{+}\theta^{-}  &  =-\theta^{-}\theta^{+},\nonumber\\
\theta^{\pm}\theta^{3}  &  =-q^{\pm2}\theta^{3}\theta^{\pm}.\nonumber
\end{align}
Using the above relations one can show that the product of two supernumbers
represented by
\begin{align}
&  f(\theta^{+},\theta^{3},\theta^{-})\\
&  =f^{\prime}+f_{+}\,\theta^{+}+f_{3}\,\theta^{3}+f_{-}\,\theta
^{-}\nonumber\\
&  +f_{+3}\,\theta^{+}\theta^{3}+f_{+-}\,\theta^{+}\theta^{-}+f_{3-\,}%
\theta^{3}\theta^{-}\nonumber\\
&  +f_{+3-}\,\theta^{+}\theta^{3}\theta^{-},\nonumber
\end{align}
now becomes
\begin{align}
&  (f\cdot g)(\theta^{+},\theta^{3},\theta^{-})\\
&  =(f\cdot g)^{\prime}+(f\cdot g)_{+}\,\theta^{+}+(f\cdot g)_{3}\,\theta
^{3}+(f\cdot g)_{-}\,\theta^{-}\nonumber\\
&  +(f\cdot g)_{+3}\,\theta^{+}\theta^{3}+(f\cdot g)_{+-}\,\theta^{+}%
\theta^{-}+(f\cdot g)_{3-}\,\theta^{3}\theta^{-}\nonumber\\
&  +(f\cdot g)_{+3-}\,\theta^{+}\theta^{3}\theta^{-},\nonumber
\end{align}
with
\begin{align}
(f\cdot g)^{\prime}  &  =f^{\prime}g^{\prime},\\
(f\cdot g)_{A}  &  =f_{A}\,g^{\prime}+f^{\prime}g_{A},\quad A\in
\{+,3,-\},\nonumber\\[0.16in]
(f\cdot g)_{+3}  &  =f_{+}\,g_{3}-q^{-2}f_{3}\,g_{+}+f^{\prime}g_{+3}%
+g^{\prime}f_{+3},\\
(f\cdot g)_{3-}  &  =f_{3}\,g_{-}-q^{-2}f_{-}\,g_{3}+f^{\prime}g_{3-}%
+g^{\prime}f_{3-},\nonumber\\
(f\cdot g)_{+-}  &  =f_{+}\,g_{-}-f_{-}\,g_{+}+\lambda f_{3}\,g_{3}\nonumber\\
&  +f^{\prime}g_{+-}+g^{\prime}f_{+-},\nonumber\\[0.16in]
(f\cdot g)_{+3-}  &  =f_{+}\,g_{3-}-q^{-2}f_{3}\,g_{+-}+q^{-2}f_{-}\,g_{+3}\\
&  +f_{+3}\,g_{-}-q^{-2}f_{+-}\,g_{3}+q^{-2}f_{3-}\,g_{+}\nonumber\\
&  +f^{\prime}g_{+3-}+g^{\prime}f_{+3-}.\nonumber
\end{align}

Next, we come to the action of symmetry generators on supernumbers. To this
end, let us notice that fermionic coordinates of three-dimensional q-deformed
Euclidean space transform under the action of $U_{q}(su_{2})$ like the
components of a vector. Thus, the commutation relations between the generators
of $U_{q}(su_{2})$ and the fermionic coordinates read
\begin{align}
L^{+}\theta^{+}  &  =\theta^{+}L^{+},\\
L^{+}\theta^{3}  &  =\theta^{3}L^{+}-q\theta^{+}\tau^{-1/2},\nonumber\\
L^{+}\theta^{-}  &  =\theta^{-}L^{+}-\theta^{3}\tau^{-1/2},\nonumber\\[0.16in]
L^{-}\theta^{+}  &  =\theta^{+}L^{-}+\theta^{3}\tau^{-1/2},\\
L^{-}\theta^{3}  &  =\theta^{3}L^{-}+q^{-1}\theta^{-}\tau^{-1/2},\nonumber\\
L^{-}\theta^{-}  &  =\theta^{-}L^{-},\nonumber\\[0.16in]
\tau^{-1/2}\theta^{\pm}  &  =q^{\pm2}\theta^{\pm}\tau^{-1/2},\\
\tau^{-1/2}\theta^{3}  &  =\theta^{3}\tau^{-1/2}.\nonumber
\end{align}
From these relations we get the representations
\begin{align}
&  L^{+}\triangleright f(\theta^{+},\theta^{3},\theta^{-})\\
&  =-qf_{3}\,\theta^{+}-f_{-}\,\theta^{3}-f_{+-}\,\theta^{+}\theta^{3}%
-qf_{3-}\,\theta^{+}\theta^{-},\nonumber\\[0.16in]
&  L^{-}\triangleright f(\theta^{+},\theta^{3},\theta^{-})\\
&  =q^{-1}f_{3}\,\theta^{-}+f_{+}\,\theta^{3}+qf_{+3}\,\theta^{+}\theta
^{-}+q^{-2}f_{+-}\,\theta^{3}\theta^{-},\nonumber\\[0.16in]
&  \tau^{-1/2}\triangleright f(\theta^{+},\theta^{3},\theta^{-})\\
&  =f(q^{2}\theta^{+},\theta^{3},q^{-2}\theta^{-}).\nonumber
\end{align}
They are related with right representations by either the transformation
rules
\begin{equation}
f(\theta^{+},\theta^{3},\theta^{-})\triangleleft L^{\pm}\overset
{+\leftrightarrow-}{\longleftrightarrow}L^{\mp}\triangleright f(\theta
^{+},\theta^{3},\theta^{-}),
\end{equation}
or
\begin{equation}
f(\theta^{+},\theta^{3},\theta^{-})\triangleleft\tau^{-1/2}=\tau
^{1/2}\triangleright f(\theta^{+},\theta^{3},\theta^{-}),
\end{equation}
The symbol $\overset{+\leftrightarrow-}{\longleftrightarrow}$ indicates the
transitions
\begin{align}
\theta^{A_{1}}\cdots\theta^{A_{n}}  &  \overset{+\leftrightarrow
-}{\longleftrightarrow}\theta^{\overline{A_{n}}}\cdots\theta^{\overline{A_{1}%
}},\\
f_{A_{1}\cdots A_{n}}  &  \overset{+\leftrightarrow-}{\longleftrightarrow
}f_{\overline{A_{n}}\cdots\overline{A_{1}}},\nonumber\\
f^{\prime}  &  \overset{+\leftrightarrow-}{\longleftrightarrow}f^{\prime
},\nonumber
\end{align}
where we have introduced indices with bar by $\bar{A}=\overline{(+,3,-)}%
=(-,3,+)$.

Now, we turn to the differentials, which have to be subject to the relations
\begin{equation}
\xi^{A}\xi^{B}=-q^{4}\hat{R}_{CD}^{AB}\,\xi^{C}\xi^{D},\quad\eta^{A}\eta
^{B}=\hat{R}_{CD}^{AB}\,\eta^{C}\eta^{D},\nonumber
\end{equation}
and
\begin{equation}
X^{A}\xi^{B}=q^{4}\hat{R}_{CD}^{AB}\,\xi^{C}X^{D},\quad\theta^{A}\eta^{B}%
=\hat{R}_{CD}^{AB}\,\eta^{C}\theta^{D}.
\end{equation}
The last two relations imply the Leibniz rules
\begin{align}
\partial_{x}^{A}X^{B}  &  =g^{AB}+(\hat{R}^{-1})_{CD}^{AB}\,X^{C}\partial
_{x}^{D},\\
\partial_{\theta}^{A}\theta^{B}  &  =g^{AB}-q^{-4}(\hat{R}^{-1})_{CD}%
^{AB}\,\theta^{C}\partial_{\theta}^{D},\nonumber
\end{align}
where $g^{AB}$ denotes the quantum metric of the three-dimensional q-deformed
Euclidean space. In complete analogy to the previous section, the relations
for the conjugated quantities follow from the above identities by applying the
substitutions%
\begin{gather}
\partial_{a}\rightarrow\hat{\partial}_{a},\quad a\rightarrow\bar{a},\quad
a\in\{\xi,\eta,X,\theta\},\\
q\rightarrow q^{-1},\quad\hat{R}\rightarrow\hat{R}^{-1}.\nonumber
\end{gather}

For the fermionic derivatives the Leibniz rules read explicitly
\begin{align}
\partial_{\theta}^{+}\theta^{+}  &  =-q^{-4}\theta^{+}\partial_{\theta}%
^{+},\label{3dimAbXanf}\\
\partial_{\theta}^{+}\theta^{3}  &  =-q^{-2}\theta^{3}\partial_{\theta}%
^{+}+q^{-2}\lambda\lambda_{+}\theta^{+}\partial_{\theta}^{3},\nonumber\\
\partial_{\theta}^{+}\theta^{-}  &  =-q-\theta^{-}\partial_{\theta}^{+}%
+q^{-1}\lambda\lambda_{+}\theta^{3}\partial_{\theta}^{3}-q^{-1}\lambda
^{2}\lambda_{+}\theta^{+}\partial_{\theta}^{-},\nonumber\\[0.16in]
\partial_{\theta}^{3}\theta^{+}  &  =-q^{-2}\theta^{+}\partial_{\theta}^{3},\\
\partial_{\theta}^{3}\theta^{3}  &  =1-q^{-2}\theta^{3}\partial_{\theta}%
^{3}+q^{-1}\lambda\lambda_{+}\theta^{+}\partial_{\theta}^{-},\nonumber\\
\partial_{\theta}^{3}\theta^{-}  &  =-q^{-2}\theta^{-}\partial_{\theta}%
^{3}+q^{-2}\lambda\lambda_{+}\theta^{3}\partial_{\theta}^{-}%
,\nonumber\\[0.16in]
\partial_{\theta}^{-}\theta^{+}  &  =-q^{-1}-\theta^{+}\partial_{\theta}%
^{-},\label{AbX3dimend}\\
\partial_{\theta}^{-}\theta^{3}  &  =-q^{-2}\theta^{3}\partial_{\theta}%
^{-},\nonumber\\
\partial_{\theta}^{-}\theta^{-}  &  =-q^{-4}\theta^{-}\partial_{\theta}%
^{-},\nonumber
\end{align}
where $\lambda_{+}\equiv q+q^{-1}.$ By the substitutions
\begin{equation}
\partial_{\theta}^{A}\rightarrow\hat{\partial}_{\theta}^{\bar{A}},\quad
\theta^{A}\rightarrow\bar{\theta}^{\bar{A}},\quad q\rightarrow q^{-1},
\end{equation}
we get the corresponding relations for the conjugated differential calculus.
In a straightforward manner, we can derive from the identities in
(\ref{3dimAbXanf})-(\ref{AbX3dimend}) the actions of the fermionic derivatives
on supernumbers. This way, we obtain
\begin{align}
&  \partial_{\theta}^{+}\triangleright f(\theta^{+},\theta^{3},\theta^{-})\\
&  =-qf_{-}+q^{-3}f_{+-}\,\theta^{+}+q^{-1}f_{3-}\,\theta^{3}-q^{-5}%
f_{+3-}\,\theta^{+}\theta^{3},\nonumber\\[0.16in]
&  \partial_{\theta}^{3}\triangleright f(\theta^{+},\theta^{3},\theta^{-})\\
&  =f_{3}-q^{-2}f_{+3}\,\theta^{+}+f_{3-}\,\theta^{-}-q^{-2}f_{+3-}%
\,\theta^{+}\theta^{-},\nonumber\\[0.16in]
&  \partial_{\theta}^{-}\triangleright f(\theta^{+},\theta^{3},\theta^{-})\\
&  =-q^{-1}f_{+}-q^{-1}f_{+3}\,\theta^{3}-q^{-1}f_{+-}\,\theta^{-}%
-q^{-1}f_{+3-}\,\theta^{3}\theta^{-}.\nonumber
\end{align}
The relationship between the different types of representations is now given
by
\begin{align}
\partial_{\theta}^{A}\triangleright f(\theta^{+},\theta^{3},\theta^{-})  &
\overset{{%
\genfrac{}{}{0pt}{}{+}{q}%
}{%
\genfrac{}{}{0pt}{}{\rightarrow}{\rightarrow}%
}{%
\genfrac{}{}{0pt}{}{-}{1/q}%
}}{\longleftrightarrow}\hat{\partial}_{\theta}^{\bar{A}}\,\bar{\triangleright
}\,f(\bar{\theta}^{-},\bar{\theta}^{3},\bar{\theta}^{+}),\\
q^{-2}f(\bar{\theta}^{+},\bar{\theta}^{3},\bar{\theta}^{-})\,\bar
{\triangleleft}\,\hat{\partial}_{\theta}^{A}\,  &  \overset{{%
\genfrac{}{}{0pt}{}{+}{q}%
}{%
\genfrac{}{}{0pt}{}{\rightarrow}{\rightarrow}%
}{%
\genfrac{}{}{0pt}{}{-}{1/q}%
}}{\longleftrightarrow}q^{2}f(\theta^{-},\theta^{3},\theta^{+})\triangleleft
\partial_{\theta}^{\bar{A}},\nonumber
\end{align}
and
\begin{align}
f(\bar{\theta}^{+},\bar{\theta}^{3},\bar{\theta}^{-})\,\bar{\triangleleft
}\,\hat{\partial}_{\theta}^{\bar{A}}  &  \overset{+\leftrightarrow
-}{\longleftrightarrow}q^{2}\partial_{\theta}^{A}\triangleright f(\theta
^{+},\theta^{3},\theta^{-}),\\
f(\theta^{-},\theta^{3},\theta^{+})\triangleleft\partial_{\theta}^{\bar{A}}
&  \overset{+\leftrightarrow-}{\longleftrightarrow}q^{-2}\hat{\partial
}_{\theta}^{A}\,\bar{\triangleright}\,f(\bar{\theta}^{-},\bar{\theta}^{3}%
,\bar{\theta}^{+}),\nonumber
\end{align}
where $\overset{{%
\genfrac{}{}{0pt}{}{+}{q}%
}{%
\genfrac{}{}{0pt}{}{\rightarrow}{\rightarrow}%
}{%
\genfrac{}{}{0pt}{}{-}{1/q}%
}}{\longleftrightarrow}$ denotes a transition given by
\begin{align}
\theta^{A_{1}}\cdots\theta^{A_{n}}  &  \overset{{%
\genfrac{}{}{0pt}{}{+}{q}%
}{%
\genfrac{}{}{0pt}{}{\rightarrow}{\rightarrow}%
}{%
\genfrac{}{}{0pt}{}{-}{1/q}%
}}{\longleftrightarrow}\theta^{\overline{A_{1}}}\cdots\theta^{\overline{A_{n}%
}},\\
f_{A_{1}\cdots A_{n}}  &  \overset{{%
\genfrac{}{}{0pt}{}{+}{q}%
}{%
\genfrac{}{}{0pt}{}{\rightarrow}{\rightarrow}%
}{%
\genfrac{}{}{0pt}{}{-}{1/q}%
}}{\longleftrightarrow}f_{\overline{A_{1}}\cdots\overline{A_{n}}},\nonumber\\
f^{\prime}  &  \overset{{%
\genfrac{}{}{0pt}{}{+}{q}%
}{%
\genfrac{}{}{0pt}{}{\rightarrow}{\rightarrow}%
}{%
\genfrac{}{}{0pt}{}{-}{1/q}%
}}{\longleftrightarrow}f^{\prime},\nonumber\\
q  &  \overset{{%
\genfrac{}{}{0pt}{}{+}{q}%
}{%
\genfrac{}{}{0pt}{}{\rightarrow}{\rightarrow}%
}{%
\genfrac{}{}{0pt}{}{-}{1/q}%
}}{\longleftrightarrow}q^{-1}.\nonumber
\end{align}

Last but not least we would like to concentrate our attention to the Hopf
structures for the various types of quantum spaces. In general we have
\begin{align}
\Delta(a^{-})  &  =a^{-}\otimes1+\Lambda(a)\tau^{-1/2}\otimes a^{-},\\
\Delta(a^{3})  &  =a^{3}\otimes1+\Lambda(a)\otimes a^{3}+\lambda\lambda
_{+}\Lambda(a)L^{+}\otimes a^{-},\nonumber\\
\Delta(a^{+})  &  =a^{+}\otimes1+\Lambda(a)\tau^{1/2}\otimes a^{+}%
+q\lambda\lambda_{+}\Lambda(a)\tau^{1/2}L^{+}\otimes a^{3}\nonumber\\
&  +q^{2}\lambda^{2}\lambda_{+}\Lambda(a)\tau^{1/2}(L^{+})^{2}\otimes
a^{-},\nonumber\\[0.16in]
S(a^{-})  &  =-\Lambda^{-1}(a)\tau^{1/2}a^{-},\\
S(a^{3})  &  =-\Lambda^{-1}(a)a^{3}+q^{2}\lambda\lambda_{+}\Lambda^{-1}%
(a)\tau^{1/2}L^{+}a^{-},\nonumber\\
S(a^{+})  &  =-\Lambda^{-1}(a)\tau^{-1/2}a^{+}+q\lambda\lambda_{+}\Lambda
^{-1}(a)L^{+}a^{3}\nonumber\\
&  -q^{4}\lambda^{2}\lambda_{+}\Lambda^{-1}(a)\tau^{1/2}(L^{+})^{2}%
a^{-},\nonumber\\[0.16in]
\varepsilon(a^{+})  &  =\varepsilon(a^{3})=\varepsilon(a^{-})=0,
\end{align}
and likewise in the conjugated case
\begin{align}
\bar{\Delta}(a^{+})  &  =a^{+}\otimes1+\Lambda^{-1}(a)\tau^{-1/2}\otimes
a^{+},\\
\bar{\Delta}(a^{3})  &  =a^{3}\otimes1+\Lambda^{-1}(a)\otimes a^{3}%
+\lambda\lambda_{+}\Lambda^{-1}(a)L^{-}\otimes a^{+},\nonumber\\
\bar{\Delta}(a^{-})  &  =a^{-}\otimes1+\Lambda^{-1}(a)\tau^{-1/2}\otimes
a^{-}+q^{-1}\lambda\lambda_{+}\Lambda^{-1}(a)\tau^{1/2}L^{-}\otimes
a^{3}\nonumber\\
&  +q^{-2}\lambda^{2}\lambda_{+}\Lambda^{-1}(a)\tau^{1/2}(L^{-})^{2}\otimes
a^{+},\nonumber\\[0.16in]
\bar{S}(a^{+})  &  =-\Lambda(a)\tau^{1/2}a^{+},\\
\bar{S}(a^{3})  &  =-\Lambda(a)a^{3}+q^{-2}\lambda\lambda_{+}\Lambda
(a)\tau^{1/2}L^{-}a^{+},\nonumber\\
\bar{S}(a^{-})  &  =-\Lambda(a)\tau^{-1/2}a^{-}+q^{-1}\lambda\lambda
_{+}\Lambda(a)L^{-}a^{3}\nonumber\\
&  -q^{-4}\lambda^{2}\lambda_{+}\Lambda(a)\tau^{1/2}(L^{-})^{2}a^{+}%
,\nonumber\\[0.16in]
\bar{\varepsilon}(a^{+})  &  =\bar{\varepsilon}(a^{3})=\bar{\varepsilon}%
(a^{-})=0,
\end{align}
where $a$ again denotes one of the following objects:
\begin{equation}
a\in\{\partial_{x},\partial_{\theta},X,\theta,\xi,\eta\}.
\end{equation}
The scaling operators have to be specified by
\begin{equation}
\Lambda(\partial_{x}^{A})=\Lambda^{1/2},\quad\Lambda(X^{A})=\Lambda
^{-1/2},\quad\Lambda(\eta^{A})=\Lambda^{1/2},
\end{equation}
and
\begin{equation}
\Lambda(\partial_{\theta}^{A})=\tilde{\Lambda},\quad\Lambda(\theta^{A}%
)=\tilde{\Lambda}^{-1},\quad\Lambda(\xi^{A})=\tilde{\Lambda}^{-1},
\end{equation}
which requires for the operators $\Lambda$ and $\tilde{\Lambda}$ to satisfy%
\begin{align}
\Lambda X^{A}  &  =q^{4}X^{A}\Lambda, & \tilde{\Lambda}X^{A}  &  =q^{2}%
X^{A}\tilde{\Lambda},\\
\Lambda\partial_{x}^{A}  &  =q^{-4}\partial_{x}^{A}\Lambda, & \tilde{\Lambda
}\partial_{x}^{A}  &  =q^{-2}\partial_{x}^{A}\tilde{\Lambda},\nonumber\\
\Lambda\xi^{A}  &  =q^{4}\xi^{A}\Lambda, & \tilde{\Lambda}\xi^{A}  &
=-q^{-2}\xi^{A}\tilde{\Lambda},\nonumber\\
\Lambda\eta^{A}  &  =q^{-4}\eta^{A}\Lambda, & \tilde{\Lambda}\eta^{A}  &
=q^{-2}\eta^{A}\tilde{\Lambda},\nonumber\\
\Lambda\theta^{A}  &  =q^{4}\theta^{A}\Lambda, & \tilde{\Lambda}\theta^{A}  &
=-q^{-2}\theta^{A}\tilde{\Lambda},\nonumber\\
\Lambda\partial_{\theta}^{A}  &  =q^{-4}\partial_{\theta}^{A}\Lambda, &
\tilde{\Lambda}\partial_{\theta}^{A}  &  =-q^{2}\partial_{\theta}^{A}%
\tilde{\Lambda}.\nonumber
\end{align}

\section{Four-dimensional Euclidean space}

The q-deformed Euclidean space in four dimensions can be treated in very much
the same way as the Euclidean space in three dimensions. Thus, we summarize
our results, only. The projector decomposition for the R-matrix is
\cite{KS97,Oca96}
\begin{equation}
\hat{R}^{-1}=q^{-1}P_{S}-qP_{A}+q^{3}P_{T}.
\end{equation}
The commutation relations among the fermionic coordinates can be written in
the general form
\begin{equation}
\theta^{i}\theta^{j}=-q\hat{R}_{kl}^{ij}\theta^{k}\theta^{l},
\end{equation}
which leads to the independent relations
\begin{align}
(\theta^{i})^{2}  &  =0,\quad i=1,\ldots,4,\\
\theta^{1}\theta^{j}  &  =-q^{-1}\theta^{j}\theta^{1},\quad j=1,2,\nonumber\\
\theta^{j}\theta^{4}  &  =-q^{-1}\theta^{4}\theta^{j},\nonumber\\
\theta^{1}\theta^{4}  &  =-\theta^{4}\theta^{1},\nonumber\\
\theta^{2}\theta^{3}  &  =-\theta^{3}\theta^{2}+\lambda\theta^{1}\theta
^{4}.\nonumber
\end{align}
For supernumbers of the form
\begin{align}
&  f(\theta^{1},\theta^{2},\theta^{3},\theta^{4})\\
&  =f^{\prime}+\sum_{i=1}^{4}f_{i}\theta^{i}+\sum_{1\leq i_{1}<i_{2}\leq
4}f_{i_{1}i_{2}}\theta^{i_{1}}\theta^{i_{2}}\nonumber\\
&  +\sum_{1\leq i_{1}<i_{2}<i_{3}\leq4}f_{i_{1}i_{2}i_{3}}\theta^{i_{1}}%
\theta^{i_{2}}\theta^{i_{3}}+f_{1234}\theta^{1}\theta^{2}\theta^{3}\theta
^{4},\nonumber
\end{align}
we can again calculate an expression for their product. Explicitly, we have
\begin{align}
&  (f\cdot g)(\theta^{1},\theta^{2},\theta^{3},\theta^{4})\\
&  =(f\cdot g)^{\prime}+\sum_{i=1}^{4}(f\cdot g)_{i}\theta^{i}+\sum_{1\leq
i_{1}<i_{2}\leq4}(f\cdot g)_{i_{1}i_{2}}\theta^{i_{1}}\theta^{i_{2}%
}\nonumber\\
&  +\sum_{1\leq i_{1}<i_{2}<i_{3}\leq4}(f\cdot g)_{i_{1}i_{2}i_{3}}%
\theta^{i_{1}}\theta^{i_{2}}\theta^{i_{3}}+(f\cdot g)_{1234}\theta^{1}%
\theta^{2}\theta^{3}\theta^{4},\nonumber
\end{align}
with
\begin{align}
(f\cdot g)^{\prime}  &  =f^{\prime}g^{\prime},\\
(f\cdot g)_{i}  &  =f_{i}g^{\prime}+f^{\prime}g_{i},\quad i=1,\ldots
,4,\nonumber\\[0.16in]
(f\cdot g)_{1j}  &  =f_{1j}g^{\prime}+f^{\prime}g_{1j}+f_{1}g_{j}-qf_{j}%
g_{1},\quad j=2,3,\\
(f\cdot g)_{j4}  &  =f_{j4}g^{\prime}+f^{\prime}g_{j4}+f_{j}g_{4}-qf_{4}%
g_{j},\nonumber\\
(f\cdot g)_{23}  &  =f_{23}g^{\prime}+f^{\prime}g_{23}+f_{2}g_{3}-f_{3}%
g_{2},\nonumber\\
(f\cdot g)_{14}  &  =f_{14}g^{\prime}+f^{\prime}g_{14}+f_{1}g_{4}-f_{4}%
g_{1}+\lambda f_{3}g_{2},\nonumber\\[0.16in]
(f\cdot g)_{123}  &  =f_{123}g^{\prime}+f^{\prime}g_{123}+f_{1}g_{23}%
-qf_{2}g_{13}+qf_{3}g_{12}\\
&  +f_{12}g_{3}-f_{13}g_{2}+q^{2}f_{23}g_{1},\nonumber\\
(f\cdot g)_{124}  &  =f_{124}g^{\prime}+f^{\prime}g_{124}+f_{1}g_{24}%
-qf_{2}g_{14}+qf_{4}g_{12}\nonumber\\
&  +f_{12}g_{4}-qf_{14}g_{2}-q\lambda f_{23}g_{2}+qf_{24}g_{1},\nonumber\\
(f\cdot g)_{134}  &  =f_{134}g^{\prime}+f^{\prime}g_{134}+f_{1}g_{34}%
+qf_{3}g_{14}-q\lambda f_{3}g_{23}+qf_{4}g_{13}\nonumber\\
&  +f_{13}\,g_{4}-qf_{14}\,g_{3}+qf_{34}\,g_{1},\nonumber\\
(f\cdot g)_{234}  &  =f_{234}\,g^{\prime}+f^{\prime}g_{234}+f_{2}%
\,g_{34}-f_{3}\,g_{24}+q^{2}f_{4}\,g_{23}\nonumber\\
&  +f_{23}\,g_{4}-qf_{24}\,g_{3}+qf_{34}\,g_{2},\nonumber\\[0.16in]
(f\cdot g)_{1234}  &  =f_{1234}\,g^{\prime}+f^{\prime}g_{1234}+f_{1}%
g_{234}-qf_{2}\,g_{134}+qf_{3}\,g_{124}\\
&  -q^{2}f_{4}g_{123}+f_{12}g_{34}-f_{13}g_{24}+q^{2}f_{14}g_{23}+q^{2}%
f_{23}g_{14}\nonumber\\
&  -q^{2}\lambda f_{23}g_{23}-q^{2}f_{24}g_{13}+q^{2}f_{34}g_{12}\nonumber\\
&  +f_{123}g_{4}-qf_{124}g_{3}+qf_{134}g_{2}-q^{2}f_{234}g_{1}.\nonumber
\end{align}

Next, we come to the commutation relations between symmetry generators of
$U_{q}(so_{4})$ (we use here the form as it was presented in Ref.
\cite{Oca96}) and fermionic coordinates:
\begin{align}
L_{1}^{+}\theta^{1}  &  =q\theta^{1}L_{1}^{+}-q^{-1}\theta^{2},\\
L_{1}^{+}\theta^{2}  &  =q^{-1}\theta^{2}L_{1}^{+},\nonumber\\
L_{1}^{+}\theta^{3}  &  =q\theta^{3}L_{1}^{+}+q^{-1}\theta^{4},\nonumber\\
L_{1}^{+}\theta^{4}  &  =q^{-1}\theta^{4}L_{1}^{+},\nonumber\\[0.16in]
L_{2}^{+}\theta^{1}  &  =q\theta^{1}L_{2}^{+}-q^{-1}\theta^{3},\\
L_{2}^{+}\theta^{2}  &  =q\theta^{2}L_{2}^{+}+q^{-1}\theta^{4},\nonumber\\
L_{2}^{+}\theta^{3}  &  =q^{-1}\theta^{3}L_{2}^{+},\nonumber\\
L_{2}^{+}\theta^{4}  &  =q^{-1}\theta^{4}L_{2}^{+},\nonumber\\[0.16in]
L_{1}^{-}\theta^{1}  &  =q\theta^{1}L_{1}^{-},\\
L_{1}^{-}\theta^{2}  &  =q^{-1}\theta^{2}L_{1}^{-}-q\theta^{1},\nonumber\\
L_{1}^{-}\theta^{3}  &  =q\theta^{3}L_{1}^{-},\nonumber\\
L_{1}^{-}\theta^{4}  &  =q^{-1}\theta^{4}L_{1}^{-}+q\theta^{3}%
,\nonumber\\[0.16in]
L_{2}^{-}\theta^{1}  &  =q\theta^{1}L_{2}^{-},\\
L_{2}^{-}\theta^{2}  &  =q\theta^{2}L_{2}^{-},\nonumber\\
L_{2}^{-}\theta^{3}  &  =q^{-1}\theta^{3}L_{2}^{-}-q\theta^{1},\nonumber\\
L_{2}^{-}\theta^{4}  &  =q^{-1}\theta^{4}L_{2}^{-}+q\theta^{2}%
,\nonumber\\[0.16in]
K_{1}\theta^{1}  &  =q^{-1}\theta^{1}K_{1},\\
K_{1}\theta^{2}  &  =q\theta^{2}K_{1},\nonumber\\
K_{1}\theta^{3}  &  =q^{-1}\theta^{3}K_{1},\nonumber\\
K_{1}\theta^{4}  &  =q\theta^{4}K_{1},\nonumber\\[0.16in]
K_{2}\theta^{1}  &  =q^{-1}\theta^{1}K_{2},\\
K_{2}\theta^{2}  &  =q^{-1}\theta^{2}K_{2},\nonumber\\
K_{2}\theta^{3}  &  =q\theta^{3}K_{2},\nonumber\\
K_{2}\theta^{4}  &  =q\theta^{4}K_{2}.\nonumber
\end{align}
With these relations it is straightforward to show that the actions of the
symmetry generators on supernumbers take the form
\begin{align}
&  L_{1}^{+}\triangleright f(\theta^{1},\theta^{2},\theta^{3},\theta^{4})\\
&  =-q^{-1}f_{1}\theta^{2}+q^{-1}f_{3}\theta^{4}\nonumber\\
&  +f_{13}\theta^{1}\theta^{4}-q^{-1}f_{13}\theta^{2}\theta^{3}\nonumber\\
&  +(q^{-2}f_{23}-q^{-1}f_{14})\theta^{2}\theta^{4}\nonumber\\
&  +q^{-1}f_{123}\theta^{1}\theta^{2}\theta^{4}-q^{-1}f_{134}\theta^{2}%
\theta^{3}\theta^{4},\nonumber\\[0.16in]
&  L_{2}^{+}\triangleright f(\theta^{1},\theta^{2},\theta^{3},\theta^{4})\\
&  =-q^{-1}f_{1}\theta^{3}+q^{-1}f_{2}\theta^{4}\nonumber\\
&  +q^{-2}f_{12}\theta^{1}\theta^{4}+q^{1}f_{12}\theta^{2}\theta
^{3}\nonumber\\
&  -(q^{-1}f_{14}+f_{23})\theta^{3}\theta^{4}\nonumber\\
&  -q^{-1}f_{123}\theta^{1}\theta^{3}\theta^{4}+q^{-1}f_{124}\theta^{2}%
\theta^{3}\theta^{4},\nonumber\\[0.16in]
&  L_{1}^{-}\triangleright f(\theta^{1},\theta^{2},\theta^{3},\theta^{4})\\
&  =-qf_{2}\theta^{1}+qf_{4}\theta^{3}\nonumber\\
&  -qf_{24}\theta^{1}\theta^{4}+f_{24}\theta^{2}\theta^{3}\nonumber\\
&  +q(qf_{14}-f_{23})\theta^{1}\theta^{3}\nonumber\\
&  +qf_{124}\theta^{1}\theta^{2}\theta^{3}-qf_{234}\theta^{1}\theta^{3}%
\theta^{4},\nonumber\\[0.16in]
&  L_{2}^{-}\triangleright f(\theta^{1},\theta^{2},\theta^{3},\theta^{4})\\
&  =-qf_{3}\theta^{1}+qf_{4}\theta^{2}\nonumber\\
&  -q^{-1}f_{34}\theta^{1}\theta^{4}-f_{34}\theta^{2}\theta^{3}\nonumber\\
&  +q^{2}(f_{14}+qf_{23})\theta^{1}\theta^{2}\nonumber\\
&  -qf_{134}\theta^{1}\theta^{2}\theta^{3}+qf_{234}\theta^{1}\theta^{2}%
\theta^{4},\nonumber
\end{align}
and
\begin{align}
K_{1}\triangleright f(\theta^{1},\theta^{2},\theta^{3},\theta^{4})  &
=f(q^{-1}\theta^{1},q\theta^{2},q^{-1}\theta^{3},q\theta^{4}),\\
K_{2}\triangleright f(\theta^{1},\theta^{2},\theta^{3},\theta^{4})  &
=f(q^{-1}\theta^{1},q^{-1}\theta^{2},q\theta^{3},q\theta^{4}).\nonumber
\end{align}
If we are interested in right representations, we can either apply the
transformation rules
\begin{equation}
f(\theta^{1},\theta^{2},\theta^{3},\theta^{4})\triangleleft L_{i}^{\pm
}\overset{i\leftrightarrow i^{\prime}}{\longleftrightarrow}q^{\mp3}L_{i}^{\mp
}\triangleright f(\theta^{1},\theta^{2},\theta^{3},\theta^{4}),
\end{equation}
or
\begin{align}
f(\theta^{1},\theta^{2},\theta^{3},\theta^{4})\triangleleft K_{1}  &
=K_{1}^{-1}\triangleright f(\theta^{1},\theta^{2},\theta^{3},\theta^{4}),\\
f(\theta^{1},\theta^{2},\theta^{3},\theta^{4})\triangleleft K_{2}  &
=K_{2}^{-1}\triangleright f(\theta^{1},\theta^{2},\theta^{3},\theta
^{4}),\nonumber
\end{align}
where $\overset{i\leftrightarrow i^{\prime}}{\longleftrightarrow}$ denotes the
transition
\begin{align}
\theta^{i_{1}}\cdots\theta^{i_{n}}  &  \overset{i\leftrightarrow i^{\prime}%
}{\longleftrightarrow}\theta^{i_{n}^{\prime}}\cdots\theta^{i_{1}^{\prime}},\\
f_{i_{1}\cdots i_{n}}  &  \overset{i\leftrightarrow i^{\prime}}%
{\longleftrightarrow}f_{i_{n}^{\prime}\cdots i_{1}^{\prime}},\nonumber\\
f^{\prime}  &  \overset{i\leftrightarrow i^{\prime}}{\longleftrightarrow
}f^{\prime},\nonumber
\end{align}
and the conjugated index is given by $i^{\prime}\equiv5-i$.

For the differentials we know that the relations \cite{CSW91}
\begin{equation}
\xi^{i}\xi^{j}=-q\hat{R}_{kl}^{ij}\,\xi^{k}\xi^{l},\quad\eta^{i}\eta
^{j}=q^{-1}\hat{R}_{kl}^{ij}\,\eta^{i}\eta^{k},
\end{equation}
and
\begin{equation}
X^{i}\xi^{j}=q\hat{R}_{kl}^{ij}\,\xi^{k}X^{l},\quad\theta^{i}\eta^{j}%
=q^{-1}\hat{R}_{kl}^{ij}\,\eta^{i}\theta^{j},
\end{equation}
hold. Using these identities we can verify that the Leibniz rules now take the
form
\begin{align}
\partial_{x}^{i}X^{j}  &  =g^{ij}+q(\hat{R}^{-1})_{kl}^{ij}\,X^{k}\partial
_{x}^{l},\\
\partial_{\theta}^{i}\theta^{j}  &  =g^{ij}-q^{-1}(\hat{R}^{-1})_{kl}%
^{ij}\,\theta^{k}\partial_{\theta}^{l},\nonumber
\end{align}
where $g^{ij}$ denotes the four-dimensional quantum space metric. Again, the
relations of the conjugated differential calculus are obtained most easily by
applying the substitutions%
\begin{gather}
\partial_{a}\rightarrow\hat{\partial}_{a},\quad a\rightarrow\bar{a},\quad
a\in\{\xi,\eta,X,\theta\},\\
q\rightarrow q^{-1},\quad\hat{R}\rightarrow\hat{R}^{-1},\nonumber
\end{gather}

Written out explicitly, the Leibniz rules become in the fermionic case
\begin{align}
\partial_{\theta}^{1}\theta^{1}  &  =-q^{-2}\theta^{1}\partial_{\theta}^{1},\\
\partial_{\theta}^{1}\theta^{2}  &  =-q^{-1}\theta^{2}\partial_{\theta}%
^{1},\nonumber\\
\partial_{\theta}^{1}\theta^{3}  &  =-q^{-1}\theta^{3}\partial_{\theta}%
^{1},\nonumber\\
\partial_{\theta}^{1}\theta^{4}  &  =q^{-1}-\theta^{4}\partial_{\theta}%
^{1},\nonumber\\[0.16in]
\partial_{\theta}^{2}\theta^{1}  &  =-q^{-1}\theta^{1}\partial_{\theta}%
^{2}+q^{-1}\lambda\theta^{2}\partial_{\theta}^{1},\\
\partial_{\theta}^{2}\theta^{2}  &  =-q^{-2}\theta^{2}\partial_{\theta}%
^{2},\nonumber\\
\partial_{\theta}^{2}\theta^{3}  &  =1-\theta^{3}\partial_{\theta}^{2}%
-\lambda\theta^{4}\partial_{\theta}^{1},\nonumber\\
\partial_{\theta}^{2}\theta^{4}  &  =-q^{-1}\theta^{4}\partial_{\theta}%
^{2},\nonumber\\[0.16in]
\partial_{\theta}^{3}\theta^{1}  &  =-q^{-1}\theta^{1}\partial_{\theta}%
^{3}+q^{-1}\lambda\theta^{3}\partial_{\theta}^{1},\\
\partial_{\theta}^{3}\theta^{2}  &  =1-\theta^{2}\partial_{\theta}^{3}%
-\lambda\theta^{4}\partial_{\theta}^{1},\nonumber\\
\partial_{\theta}^{3}\theta^{3}  &  =-q^{-2}\theta^{3}\partial_{\theta}%
^{3},\nonumber\\
\partial_{\theta}^{3}\theta^{4}  &  =-q^{-1}\theta^{4}\partial_{\theta}%
^{3},\nonumber\\[0.16in]
\partial_{\theta}^{4}\theta^{1}  &  =q-\theta^{1}\partial_{\theta}^{4}%
-\lambda(\theta^{2}\partial_{\theta}^{3}+\theta^{3}\partial_{\theta}%
^{2}+\lambda\theta^{4}\partial_{\theta}^{1}),\\
\partial_{\theta}^{4}\theta^{2}  &  =-q^{-1}\theta^{2}\partial_{\theta}%
^{4}+q^{-1}\lambda\theta^{4}\partial_{\theta}^{2},\nonumber\\
\partial_{\theta}^{4}\theta^{3}  &  =-q^{-1}\theta^{3}\partial_{\theta}%
^{4}+q^{-1}\lambda\theta^{4}\partial_{\theta}^{3},\nonumber\\
\partial_{\theta}^{4}\theta^{4}  &  =-q^{-2}\theta^{4}\partial_{\theta}%
^{4},\nonumber
\end{align}
while the substitutions
\begin{equation}
\partial_{\theta}^{i}\rightarrow\hat{\partial}_{\theta}^{i^{\prime}}%
,\quad\theta^{i}\rightarrow\bar{\theta}^{i^{\prime}},\quad q\rightarrow
q^{-1},
\end{equation}
lead to the corresponding relations for the conjugated differential calculus.
With the same reasonings already applied in the previous sections we find
\begin{align}
&  \partial_{\theta}^{1}\,\triangleright\,f(\theta^{4},\theta^{3},\theta
^{2},\theta^{1})\label{Ver4dimAbXAnf}\\
&  =q^{-1}f_{4}+q^{-1}f_{41}\theta^{1}+q^{-1}f_{42}\theta^{2}+q^{-1}%
f_{43}\theta^{3}\nonumber\\
&  +q^{-1}f_{421}\theta^{2}\theta^{1}+q^{-1}f_{432}\theta^{3}\theta^{2}%
+q^{-1}f_{431}\theta^{3}\theta^{1}\nonumber\\
&  +q^{-1}f_{4321}\theta^{3}\theta^{2}\theta^{1},\nonumber\\[0.16in]
&  \partial_{\theta}^{2}\,\triangleright\,f(\theta^{4},\theta^{3},\theta
^{2},\theta^{1})\\
&  =f_{3}+f_{31}\theta^{1}+f_{32}\theta^{2}-q^{-1}f_{43}\theta^{4}\nonumber\\
&  +f_{321}\theta^{2}\theta^{1}-q^{-1}f_{431}\theta^{4}\theta^{1}%
-q^{-1}f_{432}\theta^{4}\theta^{2}\nonumber\\
&  -q^{-1}f_{4321}\theta^{4}\theta^{2}\theta^{1},\nonumber\\[0.16in]
&  \partial_{\theta}^{3}\,\triangleright\,f(\theta^{4},\theta^{3},\theta
^{2},\theta^{1})\\
&  =f_{2}+f_{21}\theta^{1}-q^{-2}f_{32}\theta^{3}-q^{-1}f_{42}\theta
^{4}\nonumber\\
&  -q^{-2}f_{321}\theta^{3}\theta^{1}-q^{-1}f_{421}\theta^{4}\theta^{1}%
+q^{-3}f_{432}\theta^{4}\theta^{3}\nonumber\\
&  +q^{-3}f_{4321}\theta^{4}\theta^{3}\theta^{1},\nonumber\\[0.16in]
&  \partial_{\theta}^{4}\,\triangleright\,f(\theta^{4},\theta^{3},\theta
^{2},\theta^{1})\label{Ver4dimAblXEnd}\\
&  =qf_{1}-f_{21}\theta^{2}-f_{31}\theta^{3}-q^{-1}(f_{41}-\lambda
f_{32})\theta^{4}\nonumber\\
&  +q^{-1}f_{321}\theta^{3}\theta^{2}+q^{-2}f_{421}\theta^{4}\theta
^{2}\nonumber\\
&  +q^{-2}f_{431}\theta^{4}\theta^{3}+q^{-1}\lambda f_{321}\theta^{4}%
\theta^{1}\nonumber\\
&  -q^{-3}f_{4321}\theta^{4}\theta^{3}\theta^{2}.\nonumber
\end{align}
The different types of representations are linked via
\begin{align}
&  \partial_{\theta}^{i}\triangleright f(\theta^{4},\theta^{3},\theta
^{2},\theta^{1})\\
\overset{{%
\genfrac{}{}{0pt}{}{i}{q}%
}{%
\genfrac{}{}{0pt}{}{\rightarrow}{\rightarrow}%
}{%
\genfrac{}{}{0pt}{}{i^{\prime}}{1/q}%
}}{\longleftrightarrow}  &  \hat{\partial}_{\theta}^{i^{\prime}}%
\,\bar{\triangleright}\,f(\bar{\theta}^{1},\bar{\theta}^{2},\bar{\theta}%
^{3},\bar{\theta}^{4}),\nonumber\\[0.16in]
&  q^{-2}f(\bar{\theta}^{4},\bar{\theta}^{3},\bar{\theta}^{2},\bar{\theta}%
^{1})\,\bar{\triangleleft}\,\hat{\partial}_{\theta}^{i}\\
\overset{{%
\genfrac{}{}{0pt}{}{i}{q}%
}{%
\genfrac{}{}{0pt}{}{\rightarrow}{\rightarrow}%
}{%
\genfrac{}{}{0pt}{}{i^{\prime}}{1/q}%
}}{\longleftrightarrow}  &  q^{2}f(\theta^{1},\theta^{2},\theta^{3},\theta
^{4})\triangleleft\partial_{\theta}^{i^{\prime}},\nonumber
\end{align}
and
\begin{align}
&  f(\bar{\theta}^{4},\bar{\theta}^{3},\bar{\theta}^{2},\bar{\theta}%
^{1})\,\bar{\triangleleft}\,\hat{\partial}_{\theta}^{i^{\prime}}\\
\overset{i\leftrightarrow i^{\prime}}{\longleftrightarrow}  &  q^{2}%
\partial_{\theta}^{i}\triangleright f(\theta^{4},\theta^{3},\theta^{2}%
,\theta^{1}),\nonumber\\[0.16in]
&  f(\theta^{1},\theta^{2},\theta^{3},\theta^{4})\triangleleft\partial
_{\theta}^{i^{\prime}}\\
\overset{i\leftrightarrow i^{\prime}}{\longleftrightarrow}  &  q^{-2}%
\hat{\partial}_{\theta}^{i}\,\bar{\triangleright}\,f(\bar{\theta}^{1}%
,\bar{\theta}^{2},\bar{\theta}^{3},\bar{\theta}^{4}),\nonumber
\end{align}
where $\overset{{%
\genfrac{}{}{0pt}{}{i}{q}%
}{%
\genfrac{}{}{0pt}{}{\rightarrow}{\rightarrow}%
}{%
\genfrac{}{}{0pt}{}{i^{\prime}}{1/q}%
}}{\longleftrightarrow}$ stands for
\begin{align}
\theta^{i_{1}}\cdots\theta^{i_{n}}  &  \overset{{%
\genfrac{}{}{0pt}{}{i}{q}%
}{%
\genfrac{}{}{0pt}{}{\rightarrow}{\rightarrow}%
}{%
\genfrac{}{}{0pt}{}{i^{\prime}}{1/q}%
}}{\longleftrightarrow}\theta^{i_{1}^{\prime}}\cdots\theta^{i_{n}^{\prime}},\\
f_{i_{1}\cdots i_{n}}  &  \overset{{%
\genfrac{}{}{0pt}{}{i}{q}%
}{%
\genfrac{}{}{0pt}{}{\rightarrow}{\rightarrow}%
}{%
\genfrac{}{}{0pt}{}{i^{\prime}}{1/q}%
}}{\longleftrightarrow}f_{i_{1}^{\prime}\cdots i_{n}^{\prime}},\nonumber\\
f^{\prime}  &  \overset{{%
\genfrac{}{}{0pt}{}{i}{q}%
}{%
\genfrac{}{}{0pt}{}{\rightarrow}{\rightarrow}%
}{%
\genfrac{}{}{0pt}{}{i^{\prime}}{1/q}%
}}{\longleftrightarrow}f^{\prime},\nonumber\\
q  &  \overset{{%
\genfrac{}{}{0pt}{}{i}{q}%
}{%
\genfrac{}{}{0pt}{}{\rightarrow}{\rightarrow}%
}{%
\genfrac{}{}{0pt}{}{i^{\prime}}{1/q}%
}}{\longleftrightarrow}q^{-1}.\nonumber
\end{align}

Finally, we would like to present the Hopf structures for the various
four-dimensional quantum spaces. In general, we have
\begin{align}
\Delta(a^{1})  &  =a^{1}\otimes1+\Lambda(a)K_{1}^{1/2}K_{2}^{1/2}\otimes
a^{1},\label{HopUn4dim}\\
\Delta(a^{2})  &  =a^{2}\otimes1+\Lambda(a)K_{1}^{-1/2}K_{2}^{1/2}\otimes
a^{2}\nonumber\\
&  +q\lambda\Lambda(a)K_{1}^{1/2}K_{2}^{1/2}L_{1}^{+}\otimes a^{1},\nonumber\\
\Delta(a^{3})  &  =a^{3}\otimes1+\Lambda(a)K_{1}^{1/2}K_{2}^{-1/2}\otimes
a^{3}\nonumber\\
&  +q\lambda\Lambda(a)K_{1}^{1/2}K_{2}^{1/2}L_{2}^{+}\otimes a^{1},\nonumber\\
\Delta(a^{4})  &  =a^{4}\otimes1+\Lambda(a)K_{1}^{-1/2}K_{2}^{-1/2}\otimes
a^{4}\nonumber\\
&  -q^{2}\lambda^{2}\Lambda(a)K_{1}^{1/2}K_{2}^{1/2}L_{1}^{+}L_{2}^{+}\otimes
a^{1}\nonumber\\
&  -q\lambda\Lambda(a)K_{1}^{-1/2}K_{2}^{1/2}L_{2}^{+}\otimes a^{2}\nonumber\\
&  -q\lambda\Lambda(a)K_{1}^{1/2}K_{2}^{-1/2}L_{1}^{+}\otimes a^{3}%
,\nonumber\\[0.16in]
S(a^{1})  &  =-\Lambda^{-1}(a)K_{1}^{-1/2}K_{2}^{-1/2}a^{1},\\
S(a^{2})  &  =-\Lambda^{-1}(a)K_{1}^{1/2}K_{2}^{-1/2}(a^{2}-q^{2}\lambda
L_{1}^{+}a^{1}),\nonumber\\
S(a^{3})  &  =-\Lambda^{-1}(a)K_{1}^{-1/2}K_{2}^{1/2}(a^{3}-q^{2}\lambda
L_{2}^{+}a^{1}),\nonumber\\
S(a^{4})  &  =-\Lambda^{-1}(a)K_{1}^{1/2}K_{2}^{1/2}(a^{4}+q^{2}\lambda
(L_{1}^{+}a^{3}+L_{2}^{+}a^{2}))\nonumber\\
&  -q^{4}\lambda^{2}\Lambda^{-1}(a)K_{1}^{1/2}K_{2}^{1/2}L_{1}^{+}L_{2}%
^{+}a^{1},\nonumber\\[0.16in]
\varepsilon(a^{1})  &  =\varepsilon(a^{2})=\varepsilon(a^{3})=\varepsilon
(a^{4})=0,
\end{align}
and
\begin{align}
\bar{\Delta}(a^{1})  &  =a^{1}\otimes1+\Lambda^{-1}(a)K_{1}^{-1/2}K_{2}%
^{-1/2}\otimes a^{1}\label{Hop4DimKon}\\
&  -q^{-2}\lambda^{2}\Lambda^{-1}(a)K_{1}^{1/2}K_{2}^{1/2}L_{1}^{-}L_{2}%
^{-}\otimes a^{4}\nonumber\\
&  -q^{-1}\lambda\Lambda^{-1}(a)K_{1}^{1/2}K_{2}^{-1/2}L_{1}^{-}\otimes
a^{2}\nonumber\\
&  -q^{-1}\lambda\Lambda^{-1}(a)K_{1}^{-1/2}K_{2}^{1/2}L_{2}^{-}\otimes
a^{3},\nonumber\\
\bar{\Delta}(a^{2})  &  =a^{2}\otimes1+\Lambda^{-1}(a)K_{1}^{1/2}K_{2}%
^{-1/2}\otimes a^{2}\nonumber\\
&  -q^{-1}\lambda\Lambda^{-1}(a)K_{1}^{1/2}K_{2}^{1/2}L_{2}^{-}\otimes
a^{4},\nonumber\\
\bar{\Delta}(a^{3})  &  =a^{3}\otimes1+\Lambda^{-1}(a)K_{1}^{-1/2}K_{2}%
^{1/2}\otimes a^{3}\nonumber\\
&  +q^{-1}\lambda\Lambda^{-1}(a)K_{1}^{1/2}K_{2}^{1/2}L_{1}^{-}\otimes
a^{4},\nonumber\\
\bar{\Delta}(a^{4})  &  =a^{4}\otimes1+\Lambda^{-1}(a)K_{1}^{1/2}K_{2}%
^{1/2}\otimes a^{4},\nonumber\\[0.16in]
\bar{S}(a^{1})  &  =-\Lambda(a)K_{1}^{1/2}K_{2}^{1/2}(a^{1}+q^{-2}%
\lambda(L_{1}^{-}a^{2}+L_{2}^{-}a^{3}))\\
&  +q^{-4}\lambda^{2}\Lambda(a)K_{1}^{1/2}K_{2}^{1/2}L_{1}^{-}L_{2}^{-}%
a^{4},\nonumber\\
\bar{S}(a^{2})  &  =-\Lambda(a)K_{1}^{-1/2}K_{2}^{1/2}(a^{2}-q^{-2}\lambda
L_{2}^{-}a^{4}),\nonumber\\
\bar{S}(a^{3})  &  =-\Lambda(a)K_{1}^{-1/2}K_{2}^{1/2}(a^{3}-q^{-2}\lambda
L_{1}^{-}a^{4}),\nonumber\\
\bar{S}(a^{4})  &  =-\Lambda(a)K_{1}^{-1/2}K_{2}^{-1/2}a^{4}%
,\nonumber\\[0.16in]
\bar{\varepsilon}(a^{1})  &  =\bar{\varepsilon}(a^{2})=\bar{\varepsilon}%
(a^{3})=\bar{\varepsilon}(a^{4})=0,
\end{align}
where
\begin{equation}
a\in\{\partial_{x},\partial_{\theta},X,\theta,\xi,\eta\}.
\end{equation}
In order to regain relations (\ref{Ver4dimAbXAnf})-(\ref{Ver4dimAblXEnd}) and
their conjugated versions from the L-matrices determining the coproducts in
(\ref{HopUn4dim}) and (\ref{Hop4DimKon}), we have to represent the operators
$\Lambda(a)$ as
\begin{equation}
\Lambda(\partial_{x}^{i})=\Lambda^{1/2},\quad\Lambda(X^{i})=\Lambda
^{-1/2},\quad\Lambda(\eta^{i})=\Lambda^{1/2}%
\end{equation}
and
\begin{equation}
\Lambda(\partial_{\theta}^{i})=\tilde{\Lambda}^{-1},\quad\Lambda(\theta
^{i})=\tilde{\Lambda},\quad\Lambda(\xi^{i})=\tilde{\Lambda},
\end{equation}
which requires to impose on the unitary and grouplike scaling operators
$\Lambda$ and $\tilde{\Lambda}$ the commutation relations
\begin{align}
\Lambda X^{i}  &  =q^{2}X^{i}\Lambda, & \tilde{\Lambda}X^{i}  &  =q^{-1}%
X^{i}\tilde{\Lambda},\\
\Lambda\partial_{x}^{i}  &  =q^{-2}\partial_{x}^{i}\Lambda, & \tilde{\Lambda
}\partial_{x}^{i}  &  =q\partial_{x}^{i}\tilde{\Lambda},\nonumber\\
\Lambda\xi^{i}  &  =q^{2}\xi^{i}\Lambda, & \tilde{\Lambda}\xi^{i}  &
=-q\xi^{i}\tilde{\Lambda},\nonumber\\
\Lambda\eta^{i}  &  =q^{2}\eta^{i}\Lambda, & \tilde{\Lambda}\eta^{i}  &
=q\eta^{i}\tilde{\Lambda},\nonumber\\
\Lambda\theta^{i}  &  =q^{2}\theta^{i}\Lambda, & \tilde{\Lambda}\theta^{i}  &
=-q\theta^{i}\tilde{\Lambda},\nonumber\\
\Lambda\partial_{\theta}^{i}  &  =q^{-2}\partial_{\theta}^{i}\Lambda, &
\tilde{\Lambda}\partial_{\theta}^{i}  &  =-q^{-1}\partial_{\theta}^{i}%
\tilde{\Lambda}.\nonumber
\end{align}

\section{Minkowski space}

In this section we would like to deal with q-deformed Minkowski space
\cite{CSSW90,SWZ91,LWW97,Maj91,OSWZ92} which from a physical point of view is
the most interesting case in this article (for other deformations of spacetime
and their related symmetries we refer to
\cite{Lu92,Cas93,Dov94,ChDe95,ChKu04,Koch04}). We follow the same line of
arguments as in the previous sections. The R-matrix now obeys the
decomposition \cite{LSW94}
\begin{equation}
\hat{R}_{II}=q^{-2}P_{S}-P_{A}+q^{2}P_{T}.
\end{equation}
The relations for the fermionic coordinates are completely determined by
\begin{equation}
\theta^{i}\theta^{j}=-(\hat{R}_{II})_{kl}^{ij}\theta^{k}\theta^{l},
\end{equation}
from which we obtain as independent relations
\begin{align}
(\theta^{\mu})^{2} &  =0,\quad\mu\in\{+,-,0\}\\
\theta^{3}\theta^{\pm} &  =-q^{\mp2}\theta^{\pm}\theta^{3},\nonumber\\
\theta^{3}\theta^{3} &  =\lambda\theta^{+}\theta^{-},\nonumber\\
\theta^{+}\theta^{-} &  =-\theta^{-}\theta^{+},\nonumber\\
\theta^{\pm}\theta^{0}+\theta^{0}\theta^{\pm} &  =\pm q^{\mp1}\lambda
\theta^{\pm}\theta^{3},\nonumber\\
\theta^{0}\theta^{3}+\theta^{3}\theta^{0} &  =\lambda\theta^{+}\theta
^{-}.\nonumber
\end{align}
Instead of dealing with the coordinate $\theta^{3}$ or $\theta^{0}$ it is
often more convenient to work with the light-cone coordinate $\theta
^{3/0}=\theta^{3}-\theta^{0}$, for which we have the additional relations
\begin{align}
(\theta^{3/0})^{2} &  =0,\\
\theta^{\pm}\theta^{3/0} &  =-\theta^{3/0}\theta^{\pm},\nonumber\\
\theta^{0}\theta^{3/0}+\theta^{3/0}\theta^{0} &  =-\lambda\theta^{+}\theta
^{-},\nonumber\\
\theta^{\pm}\theta^{0}+q^{\pm2}\theta^{0}\theta^{\pm} &  =\pm q^{\pm1}%
\lambda\theta^{\pm}\theta^{3/0},\nonumber\\
\theta^{3}\theta^{3/0}+\theta^{3/0}\theta^{3} &  =-\lambda\theta^{+}\theta
^{-}.\nonumber
\end{align}
The product of two supernumbers of the form
\begin{align}
&  f(\theta^{+},\theta^{3},\theta^{0},\theta^{-})\\
&  =f^{\prime}+f_{+}\theta^{+}+f_{0}\theta^{0}+f_{3}\theta^{3}+f_{-}\theta
^{-}\nonumber\\
&  +f_{+3}\theta^{+}\theta^{3}+f_{+0}\theta^{+}\theta^{0}+f_{+-}\theta
^{+}\theta^{-}\nonumber\\
&  +f_{30}\theta^{3}\theta^{0}+f_{3-}\theta^{3}\theta^{-}+f_{0-}\theta
^{0}\theta^{-}\nonumber\\
&  +f_{+30}\theta^{+}\theta^{3}\theta^{0}+f_{+3-}\theta^{+}\theta^{3}%
\theta^{-}+f_{+0-}\theta^{+}\theta^{0}\theta^{-}\nonumber\\
&  +f_{30-}\theta^{3}\theta^{0}\theta^{-}+f_{+30-}\theta^{+}\theta^{3}%
\theta^{0}\theta^{-}\nonumber
\end{align}
now becomes
\begin{align}
&  (f\cdot g)(\theta^{+},\theta^{3},\theta^{0},\theta^{-})\\
&  =(f\cdot g)^{\prime}+(f\cdot g)_{+}\theta^{+}+(f\cdot g)_{0}\theta
^{0}\nonumber\\
&  +(f\cdot g)_{3}\theta^{3}+(f\cdot g)_{-}\theta^{-}\nonumber\\
&  +(f\cdot g)_{+3}\theta^{+}\theta^{3}+(f\cdot g)_{+0}\theta^{+}\theta
^{0}+(f\cdot g)_{+-}\theta^{+}\theta^{-}\nonumber\\
&  +(f\cdot g)_{30}\theta^{3}\theta^{0}+(f\cdot g)_{3-}\theta^{3}\theta
^{-}+(f\cdot g)_{0-}\theta^{0}\theta^{-}\nonumber\\
&  +(f\cdot g)_{+30}\theta^{+}\theta^{3}\theta^{0}+(f\cdot g)_{+3-}\theta
^{+}\theta^{3}\theta^{-}\nonumber\\
&  +(f\cdot g)_{+0-}\theta^{+}\theta^{0}\theta^{-}+(f\cdot g)_{30-}\theta
^{3}\theta^{0}\theta^{-}\nonumber\\
&  +(f\cdot g)_{+30-}\theta^{+}\theta^{3}\theta^{0}\theta^{-},\nonumber
\end{align}
with
\begin{align}
(f\cdot g)^{\prime} &  =f^{\prime}g^{\prime},\\
(f\cdot g)_{\mu} &  =f_{\mu}g^{\prime}+f^{\prime}g_{\mu},\quad\mu
\in\{+,3,0,-\},\nonumber\\[0.16in]
(f\cdot g)_{+0} &  =f_{+0}g^{\prime}+f^{\prime}g_{+0}+f_{+}g_{0}-f_{0}g_{+},\\
(f\cdot g)_{30} &  =f_{30}g^{\prime}+f^{\prime}g_{30}+f_{3}g_{0}-f_{0}%
g_{3},\nonumber\\
(f\cdot g)_{0-} &  =f_{0-}g^{\prime}+f^{\prime}g_{0-}+f_{0}g_{-}-f_{-}%
g_{0},\nonumber\\
(f\cdot g)_{+-} &  =f_{+-}g^{\prime}+f^{\prime}g_{+-}+f_{+}g_{-}-f_{-}%
g_{+}+\lambda f_{3}g_{3}-\lambda f_{0}g_{3},\nonumber\\
(f\cdot g)_{+3} &  =f_{+3}g^{\prime}+f^{\prime}g_{+3}+f_{+}g_{3}-q^{-2}%
f_{3}g_{+}-q^{-1}\lambda f_{0}g_{+},\nonumber\\
(f\cdot g)_{3-} &  =f_{3-}g^{\prime}+f^{\prime}g_{3-}+f_{3}g_{-}-q^{-2}%
f_{-}g_{3}-q^{-1}\lambda f_{-}g_{0},\nonumber\\[0.16in]
(f\cdot g)_{+30} &  =f_{+30}g^{\prime}+f^{\prime}g_{30}+f_{+}g_{30}%
-q^{-2}f_{3}g_{+0}\\
&  +f_{0}g_{+3}-q^{-1}\lambda f_{0}g_{+0}\nonumber\\
&  +f_{+3}g_{0}-f_{+0}g_{3}+q^{-2}f_{30}g_{+},\nonumber\\
(f\cdot g)_{30-} &  =f_{30-}g^{\prime}+f^{\prime}g_{30-}+f_{3}g_{0-}%
-f_{0}g_{3-}+q^{-2}f_{-}g_{30}\nonumber\\
&  +f_{30}g_{-}-f_{3-}g_{0}+q^{-2}f_{0-}g_{3}+q^{-1}\lambda f_{0-}%
g_{0},\nonumber\\
(f\cdot g)_{+0-} &  =f_{+0-}g^{\prime}+f^{\prime}g_{+0-}+f_{+}g_{0-}%
-f_{0}g_{+-}-\lambda f_{3}g_{30}\nonumber\\
&  +f_{-}g_{+0}-\lambda f_{0}g_{30}+f_{+0}g_{-}\nonumber\\
&  -f_{+-}g_{0}+f_{0-}g_{+}+\lambda f_{30}g_{3},\nonumber\\
(f\cdot g)_{+3-} &  =f_{+3-}g^{\prime}+f^{\prime}g_{+3-}+q^{-1}\lambda
f_{+}g_{3-}-q^{-2}f_{-}g_{+3}\nonumber\\
&  -q^{-2}f_{3}g_{+-}-q^{-1}\lambda f_{0}g_{+-}-q^{-1}\lambda f_{-}%
g_{+0}\nonumber\\
&  -q^{-1}\lambda^{2}f_{3}g_{30}-q^{-1}\lambda^{2}f_{0}g_{30}\nonumber\\
&  +f_{+3}g_{-}-q^{-2}f_{+-}g_{3}+q^{-2}f_{3-}g_{+}\nonumber\\
&  -q^{-2}f_{+-}g_{3}-q^{-1}\lambda f_{+-}g_{0}+q^{-1}\lambda f_{0-}%
g_{+}\nonumber\\
&  +q^{-1}\lambda(\lambda-q^{-1})f_{30}g_{3},\nonumber\\[0.16in]
(f\cdot g)_{+30-} &  =f_{+30-}g^{\prime}+f^{\prime}g_{+30-}+f_{+}%
g_{30-}-q^{-2}f_{3}g_{+0-}+f_{0}g_{+3-}\\
&  -q^{-2}f_{-}g_{+30-}-q^{-1}\lambda f_{0}g_{+0-}\nonumber\\
&  +f_{+3}g_{0-}-f_{+0}g_{3-}+q^{-2}f_{+-}g_{30}+q^{-2}f_{30}g_{+-}\nonumber\\
&  -q^{-2}f_{3-}g_{+0}+q^{-2}f_{0-}g_{+3}+q^{-2}\lambda f_{30}g_{30}%
\nonumber\\
&  +f_{+30}g_{-}-f_{+3-}g_{0}+q^{-2}f_{+0-}g_{3}\nonumber\\
&  -q^{-2}f_{30-}g_{+}+q^{-1}\lambda f_{+0-}g_{0}.\nonumber
\end{align}

Next, we turn to the commutation relations between generators of q-deformed
Lorentz algebra (for its definition see Refs. \cite{SWZ91,OSWZ92}) and
fermionic coordinates. Explicitly, they read
\begin{align}
T^{+}\theta^{0}  &  =\theta^{0}T^{+},\\
T^{+}\theta^{3/0}  &  =\theta^{3/0}T^{+}+q^{-3/2}\lambda_{+}^{1/2}\theta
^{+},\nonumber\\
T^{+}\theta^{+}  &  =q^{-2}\theta^{+}T^{+},\nonumber\\
T^{+}\theta^{-}  &  =q^{2}\theta^{-}T^{+}+q^{-1/2}\lambda_{+}^{1/2}\theta
^{3},\nonumber\\[0.16in]
T^{-}\theta^{0}  &  =\theta^{0}T^{-},\\
T^{-}\theta^{3/0}  &  =\theta^{3/0}T^{-}+q^{3/2}\lambda_{+}^{1/2}\theta
^{-},\nonumber\\
T^{-}\theta^{-}  &  =q^{2}\theta^{-}T^{-},\nonumber\\
T^{-}\theta^{+}  &  =q^{-2}\theta^{+}T^{-}+q^{1/2}\lambda_{+}^{1/2}\theta
^{3},\nonumber\\[0.16in]
\tau^{3}\theta^{0}  &  =\theta^{0}\tau^{3},\\
\tau^{3}\theta^{3/0}  &  =\theta^{3/0}\tau^{3},\nonumber\\
\tau^{3}\theta^{+}  &  =q^{-4}\theta^{+}\tau^{3},\nonumber\\
\tau^{3}\theta^{-}  &  =q^{4}\theta^{-}\tau^{3},\nonumber\\[0.16in]
T^{2}\theta^{3/0}  &  =q^{-1}\theta^{3/0}T^{2},\\
T^{2}\theta^{+}  &  =q\theta^{+}T^{2},\nonumber\\
T^{2}\theta^{-}  &  =q^{-1}\theta^{-}T^{2}+q^{-3/2}\lambda_{+}^{-1/2}%
\theta^{3/0}\tau^{1},\nonumber\\
T^{2}\theta^{3}  &  =q\theta^{3}T^{2}-q\lambda_{+}^{-1}\lambda\theta
^{3/0}T^{2}+q^{-1/2}\lambda_{+}^{-1/2}\theta^{+}\tau^{1},\nonumber\\[0.16in]
S^{1}\theta^{3/0}  &  =q\theta^{3/0}S^{1},\\
S^{1}\theta^{-}  &  =q\theta^{-}S^{1},\nonumber\\
S^{1}\theta^{+}  &  =q^{-1}\theta^{+}S^{1}-q^{-1/2}\lambda_{+}^{-1/2}%
\theta^{3/0}\sigma^{2},\nonumber\\
S^{1}\theta^{3}  &  =q^{-1}\theta^{3}S^{1}+q^{-1}\lambda_{+}^{-1}\lambda
\theta^{3/0}S^{1}-q^{1/2}\lambda_{+}^{-1/2}\theta^{-}\sigma^{2}%
,\nonumber\\[0.16in]
\tau^{1}\theta^{3/0}  &  =q\theta^{3/0}\tau^{1},\\
\tau^{1}\theta^{-}  &  =q^{-1}\theta^{-}\tau^{1},\nonumber\\
\tau^{1}\theta^{+}  &  =q\theta^{+}\tau^{1}-q^{3/2}\lambda_{+}^{-1/2}%
\lambda^{2}\theta^{3/0}T^{2},\nonumber\\
\tau^{1}\theta^{3}  &  =q^{-1}\theta^{3}\tau^{1}+q^{-1}\lambda_{+}^{-1}%
\lambda\theta^{3/0}\tau^{1}-q^{1/2}\lambda_{+}^{-1/2}\lambda^{2}\theta
^{-}T^{2},\nonumber\\[0.16in]
\sigma^{2}\theta^{3/0}  &  =q^{-1}\theta^{3/0}\sigma^{2},\\
\sigma^{2}\theta^{+}  &  =q^{-1}\theta^{+}\sigma^{2},\nonumber\\
\sigma^{2}\theta^{-}  &  =q\theta^{-}\sigma^{2}+q^{1/2}\lambda_{+}%
^{-1/2}\lambda^{2}\theta^{3/0}S^{1},\nonumber\\
\sigma^{2}\theta^{3}  &  =q\theta^{3}\sigma^{2}-q\lambda_{+}^{-1}\lambda
\theta^{3/0}\sigma^{2}+q^{-1/2}\lambda_{+}^{-1/2}\lambda^{2}\theta^{+}%
S^{1}.\nonumber
\end{align}
The generators $T^{+},$ $T^{-},$ and $\tau^{3}$ span a $U_{q}(su_{2}%
)$-subalgebra of the q-deformed Lorentz algebra. With the above relations on
hand we find for its generators the following actions on supernumbers:
\begin{align}
&  \tau^{3}\triangleright f(\theta^{+},\theta^{3},\theta^{0},\theta^{-})\\
&  =f(q^{-4}\theta^{+},\theta^{3},\theta^{0},q^{4}\theta^{-}%
),\nonumber\\[0.16in]
&  T^{-}\triangleright f(\theta^{+},\theta^{3},\theta^{0},\theta^{-})\\
&  =q^{3/2}\lambda_{+}^{1/2}f_{3}\theta^{-}+q^{3/2}\lambda_{+}^{1/2}%
f_{+3}\theta^{+}\theta^{-}\nonumber\\
&  +q^{1/2}\lambda_{+}^{1/2}f_{+0}\theta^{3}\theta^{0}-q^{3/2}\lambda
_{+}^{1/2}f_{30}\theta^{0}\theta^{-}\nonumber\\
&  +q^{1/2}\lambda_{+}^{1/2}(f_{+-}+\lambda f_{30})\theta^{3}\theta
^{-}\nonumber\\
&  -q^{3/2}\lambda_{+}^{1/2}f_{3+0}\theta^{+}\theta^{0}\theta^{-}%
+q^{1/2}\lambda_{+}^{1/2}f_{+0-}\theta^{3}\theta^{0}\theta^{-}\nonumber\\
&  +q^{1/2}\lambda\lambda_{+}^{1/2}f_{+30}\theta^{+}\theta^{3}\theta
^{-},\nonumber\\[0.16in]
&  T^{+}\triangleright f(\theta^{+},\theta^{3},\theta^{0},\theta^{-})\\
&  =q^{-3/2}\lambda_{+}^{1/2}f_{3}\theta^{+}-q^{-1/2}\lambda_{+}^{1/2}%
f_{0-}\theta^{3}\theta^{0}\nonumber\\
&  +q^{-3/2}\lambda_{+}^{1/2}f_{30}\theta^{+}\theta^{0}+q^{-5/2}\lambda
_{+}^{1/2}f_{+-}\theta^{+}\theta^{3}\nonumber\\
&  +\lambda_{+}^{1/2}(q^{1/2}f_{3-}+q^{-1/2}\lambda f_{0-})\theta^{+}%
\theta^{-}\nonumber\\
&  -q^{-5/2}\lambda_{+}^{1/2}f_{+0-}\theta^{+}\theta^{3}\theta^{0}%
+q^{1/2}\lambda_{+}^{1/2}f_{30-}\theta^{+}\theta^{0}\theta^{-}\nonumber\\
&  -q^{-1/2}\lambda\lambda_{+}^{1/2}f_{30-}\theta^{+}\theta^{3}\theta
^{-}.\nonumber
\end{align}
Right representations are obtained most easily by either applying the
transformations
\begin{align}
&  f(\theta^{+},\theta^{0},\theta^{3},\theta^{-})\triangleleft T^{\pm}\\
\overset{+\leftrightarrow-}{\longleftrightarrow}  &  -q^{\mp3}T^{\mp
}\triangleright f(\theta^{+},\theta^{3},\theta^{0},\theta^{-})\nonumber
\end{align}
or the identity
\begin{equation}
f(\theta^{+},\theta^{0},\theta^{3},\theta^{-})\triangleleft\tau^{3}%
=f(q^{4}\theta^{+},\theta^{0},\theta^{3},q^{-4}\theta^{-}),
\end{equation}
where
\begin{align}
\theta^{\mu_{1}}\cdots\theta^{\mu_{n}}  &  \overset{+\leftrightarrow
-}{\longleftrightarrow}\theta^{\overline{\mu_{n}}}\cdots\theta^{\overline
{\mu_{1}}},\\
f_{\mu_{1}\cdots\mu_{n}}  &  \overset{+\leftrightarrow-}{\longleftrightarrow
}f_{\overline{\mu_{n}}\cdots\overline{\mu_{1}}},\nonumber\\
f^{\prime}  &  \overset{+\leftrightarrow-}{\longleftrightarrow}f^{\prime
},\nonumber
\end{align}
with the conjugated index now defined by
\begin{equation}
\bar{\mu}=\overline{(+,3,3/0,0,-)}=(-,3,3/0,0,+).
\end{equation}
For the remaining generators we have
\begin{align}
&  \sigma^{2}\triangleright f(\theta^{+},\theta^{3},\theta^{3/0},\theta^{-})\\
&  =q^{-1}f_{+}\theta^{+}+qf_{-}\theta^{-}+qf_{3}\theta^{3}\nonumber\\
&  +(q^{-1}f_{3/0}-q\lambda_{+}^{-1}\lambda f_{3})\theta^{3/0}\nonumber\\
&  +f_{+3}\theta^{+}\theta^{3}+f_{+-}\theta^{+}\theta^{-}+q^{2}f_{3-}%
\theta^{3}\theta^{-}\nonumber\\
&  +f_{3,3/0}\theta^{3}\theta^{3/0}+(f_{3/0,-}-q^{2}\lambda\lambda_{+}%
^{-1}f_{3-})\theta^{3/0}\theta^{-}\nonumber\\
&  +(q^{-2}f_{+,3/0}-\lambda\lambda_{+}^{-1}f_{+3})\theta^{+}\theta
^{3/0}\nonumber\\
&  +q^{-1}f_{+3,3/0}\theta^{+}\theta^{3}\theta^{3/0}+qf_{+3-}\theta^{+}%
\theta^{3}\theta^{-}+qf_{3,3/0,-}\theta^{3}\theta^{3/0}\theta^{-}\nonumber\\
&  +(q^{-1}f_{+,3/0,-}-q\lambda\lambda_{+}^{-1}f_{+3-})\theta^{+}\theta
^{3/0}\theta^{-}\nonumber\\
&  +f_{+3,3/0,-}\theta^{+}\theta^{3}\theta^{3/0}\theta^{-},\nonumber\\[0.16in]
&  \tau^{1}\triangleright f(\theta^{+},\theta^{3},\theta^{3/0},\theta^{-})\\
&  =q^{-1}f_{-}\theta^{-}+qf_{+}\theta^{+}+q^{-1}f_{3}\theta^{3}\nonumber\\
&  +(qf_{3/0}+q^{-1}\lambda_{+}^{-1}\lambda f_{3})\theta^{3/0}\nonumber\\
&  +f_{+3}\theta^{+}\theta^{3}+f_{+-}\theta^{+}\theta^{-}+q^{-2}f_{3-}%
\theta^{3}\theta^{-}\nonumber\\
&  +f_{3,3/0}\theta^{3}\theta^{3/0}+q^{2}(f_{+3/0}+\lambda\lambda_{+}%
^{-1}f_{+3})\theta^{+}\theta^{3/0}\nonumber\\
&  +(f_{3/0,-}+\lambda\lambda_{+}^{-1}f_{3-})\theta^{3/0}\theta^{-}\nonumber\\
&  +qf_{+3,3/0}\theta^{+}\theta^{3}\theta^{3/0}+q^{-1}f_{+3-}\theta^{+}%
\theta^{3}\theta^{-}+q^{-1}f_{3,3/0,-}\theta^{3}\theta^{3/0}\theta
^{-}\nonumber\\
&  +(qf_{+,3/0,-}+\lambda\lambda_{+}^{-1}(2(\lambda+q)-q^{2})f_{+3-}%
)\theta^{+}\theta^{3/0}\theta^{-}\nonumber\\
&  +f_{+3,3/0,-}\theta^{+}\theta^{3}\theta^{3/0}\theta^{-},\nonumber\\[0.16in]
&  S^{1}\triangleright f(\theta^{+},\theta^{3},\theta^{3/0},\theta^{-})\\
&  =-q^{-1/2}\lambda_{+}^{-1/2}f_{+}\theta^{3/0}-q^{1/2}\lambda_{+}%
^{-1/2}f_{3}\theta^{-}\nonumber\\
&  +q^{1/2}\lambda_{+}^{-1/2}f_{+3}\theta^{3}\theta^{3/0}+q^{-1/2}\lambda
_{+}^{-1/2}(q\lambda-1)f_{+3}\theta^{+}\theta^{-}\nonumber\\
&  +\lambda_{+}^{-1/2}(q^{-1/2}f_{3,3/0}-q^{1/2}f_{+-})\theta^{3/0}\theta
^{-}\nonumber\\
&  +q^{3/2}\lambda_{+}^{-1/2}f_{+3-}\theta^{3}\theta^{3/0}\theta
^{-}\nonumber\\
&  +q^{-3/2}\lambda_{+}^{-1/2}(1-q\lambda)f_{+3,3/0}\theta^{+}\theta
^{3/0}\theta^{-},\nonumber\\[0.16in]
&  T^{2}\triangleright f(\theta^{+},\theta^{3},\theta^{3/0},\theta^{-})\\
&  =q^{-3/2}\lambda_{+}^{-1/2}f_{-}\theta^{3/0}+q^{-1/2}\lambda_{+}%
^{-1/2}f_{3}\theta^{+}\nonumber\\
&  +q^{-1/2}\lambda_{+}^{-1/2}f_{3-}\theta^{3}\theta^{3/0}+q^{-3/2}\lambda
_{+}^{-1/2}f_{3-}\theta^{+}\theta^{-}\nonumber\\
&  +\lambda_{+}^{-1/2}(q^{1/2}f_{3,3/0}+q^{-1/2}f_{+-})\theta^{+}\theta
^{3/0}\nonumber\\
&  +q^{1/2}\lambda_{+}^{-1/2}f_{+3-}\theta^{+}\theta^{3}\theta^{3/0}%
\nonumber\\
&  +q^{-1/2}\lambda_{+}^{-1/2}f_{3,3/0,-}\theta^{+}\theta^{3/0}\theta
^{-}.\nonumber
\end{align}
The easiest way to derive the corresponding right representations is to use
the identity
\begin{equation}
S^{-1}(h)\triangleright f=f\triangleleft h
\end{equation}
together with \cite{OSWZ92}
\begin{align}
S^{-1}(T^{2})  &  =-q^{-2}T^{2}(\tau^{3})^{1/2},\\
S^{-1}(S^{1})  &  =-S^{1}(\tau^{3})^{-1/2},\nonumber\\
S^{-1}(\tau^{1})  &  =\sigma^{2},\nonumber\\
S^{-1}(\sigma^{2})  &  =\tau^{1}.\nonumber
\end{align}

Now, let us consider the differentials, which obey the commutation relations
\cite{OSWZ92}
\begin{equation}
\xi^{\mu}\xi^{\nu}=-(\hat{R}_{II})_{\rho\sigma}^{\mu\nu}\,\xi^{\rho}%
\xi^{\sigma},\quad\eta^{\mu}\eta^{\nu}=q^{2}(\hat{R}_{II})_{\rho\sigma}%
^{\mu\nu}\,\eta^{\rho}\eta^{\sigma}, \label{DDVerMinkow}%
\end{equation}
and
\begin{equation}
X^{\mu}\xi^{\nu}=(\hat{R}_{II})_{\rho\sigma}^{\mu\nu}\,\xi^{\rho}X^{\sigma
},\quad\theta^{\mu}\eta^{\nu}=q^{2}(\hat{R}_{II})_{\rho\sigma}^{\mu\nu}%
\,\eta^{\rho}\theta^{\sigma}. \label{KDRelMinkow}%
\end{equation}
The Leibniz rules being compatible with the identities in (\ref{KDRelMinkow})
read
\begin{align}
\partial_{x}^{\mu}X^{\nu}  &  =\eta^{\mu\nu}+q^{-2}(\hat{R}_{II}^{-1}%
)_{\rho\sigma}^{\mu\nu}\,X^{\rho}\partial_{x}^{\sigma},\label{LeibMinferm}\\
\partial_{\theta}^{\mu}\theta^{\nu}  &  =\eta^{\mu\nu}-q^{-1}(\hat{R}%
_{II}^{-1})_{\rho\sigma}^{\mu\nu}\,\theta^{\rho}\partial_{\theta}^{\sigma
},\nonumber
\end{align}
where $\eta^{\mu\nu}$ stands for the metric of q-deformed Minkowski space.
With the substitutions
\begin{gather}
\partial_{a}\rightarrow\hat{\partial}_{a},\quad a\rightarrow\bar{a},\quad
a\in\{\xi,\eta,X,\theta\},\label{SubConMin}\\
q\rightarrow q^{-1},\quad\hat{R}\rightarrow\hat{R}^{-1},\nonumber
\end{gather}
the formulae in (\ref{DDVerMinkow})-(\ref{LeibMinferm}) transform into those
of the conjugated calculus.

As in the previous sections, we would like to write down the Leibniz rules for
the fermionic derivatives, explicitly. In this way we have
\begin{align}
\partial_{\theta}^{3/0}\theta^{3/0}  &  =-q^{2}\theta^{3/0}\partial_{\theta
}^{3/0},\label{MinLeibRulAnf}\\
\partial_{\theta}^{3/0}\theta^{+}  &  =-q^{2}\theta^{+}\partial_{\theta}%
^{3/0}-q\lambda\theta^{3/0}\partial_{\theta}^{+},\nonumber\\
\partial_{\theta}^{3/0}\theta^{3}  &  =1-\theta^{3}\partial_{\theta}%
^{3/0}-\lambda\lambda_{+}^{-1}\theta^{3/0}\partial_{\theta}^{3/0}%
-\lambda\theta^{-}\partial_{\theta}^{+},\nonumber\\
\partial_{\theta}^{3/0}\theta^{-}  &  =-\theta^{-}\partial_{\theta}%
^{3/0},\nonumber\\[0.16in]
\partial_{\theta}^{+}\theta^{3/0}  &  =-\theta^{3/0}\partial_{\theta}^{+},\\
\partial_{\theta}^{+}\theta^{+}  &  =-q^{2}\theta^{+}\partial_{\theta}%
^{+},\nonumber\\
\partial_{\theta}^{+}\theta^{3}  &  =-q^{2}\theta^{3}\partial_{\theta}%
^{+}+q^{2}\lambda\lambda_{+}^{-1}\theta^{3/0}\partial_{\theta}^{+}%
+q^{2}\lambda\lambda_{+}^{-1}\theta^{+}\partial_{\theta}^{3/0},\nonumber\\
\partial_{\theta}^{+}\theta^{-}  &  =-q-\theta^{-}\partial_{\theta}%
^{+}+q\lambda\lambda_{+}^{-1}\theta^{3/0}\partial_{\theta}^{3/0}%
,\nonumber\\[0.16in]
\partial_{\theta}^{-}\theta^{3/0}  &  =-q^{2}\theta^{3/0}\partial_{\theta}%
^{-}-q\lambda\theta^{-}\partial_{\theta}^{3/0},\\
\partial_{\theta}^{-}\theta^{+}  &  =-q^{-1}-\theta^{+}\partial_{\theta}%
^{-}-\lambda\theta^{3/0}\partial_{\theta}^{0}-\lambda^{2}\theta^{-}%
\partial_{\theta}^{+}\nonumber\\
&  -\lambda\theta^{3}\partial_{\theta}^{3/0}-q\lambda\lambda_{+}^{-1}%
\theta^{3/0}\partial_{\theta}^{3/0},\nonumber\\
\partial_{\theta}^{-}\theta^{3}  &  =-\theta^{3}\partial_{\theta}^{-}%
-\lambda\lambda_{+}^{-1}\theta^{3/0}\partial_{\theta}^{-}-q\lambda\theta
^{-}\partial_{\theta}^{0}\nonumber\\
&  -\lambda\lambda_{+}^{-1}(q^{2}+2)\theta^{-}\partial_{\theta}^{3/0}%
,\nonumber\\
\partial_{\theta}^{-}\theta^{-}  &  =-q^{2}\theta^{-}\partial_{\theta}%
^{-},\nonumber\\[0.16in]
\partial_{\theta}^{0}\theta^{3/0}  &  =1-\theta^{3/0}\partial_{\theta}%
^{0}-\lambda\theta^{-}\partial_{\theta}^{+}+q^{2}\lambda\lambda_{+}^{-1}%
\theta^{3/0}\partial_{\theta}^{3/0},\label{MinLeibRulEnd}\\
\partial_{\theta}^{0}\theta^{+}  &  =-\theta^{+}\partial_{\theta}^{0}%
-q\lambda\theta^{3}\partial_{\theta}^{+}+q^{2}\lambda\lambda_{+}^{-1}%
\theta^{3/0}\partial_{\theta}^{+}+q^{2}\lambda\lambda_{+}^{-1}\theta
^{+}\partial_{\theta}^{3/0},\nonumber\\
\partial_{\theta}^{0}\theta^{3}  &  =-q^{2}\theta^{3}\partial_{\theta}%
^{0}+q^{2}\lambda\lambda_{+}^{-1}\theta^{3/0}\partial_{\theta}^{0}%
+q\lambda\lambda_{+}^{-1}\theta^{+}\partial_{\theta}^{-}-q\lambda\lambda
_{+}^{-1}\theta^{-}\partial_{\theta}^{+}\nonumber\\
&  +\lambda\lambda_{+}^{-1}\theta^{3}\partial_{\theta}^{3/0}+q^{2}%
\lambda\lambda_{+}^{-1}\theta^{3/0}\partial_{\theta}^{3/0},\nonumber\\
\partial_{\theta}^{0}\theta^{-}  &  =-q^{2}\theta^{-}\partial_{\theta}%
^{0}-\lambda\lambda_{+}^{-1}\theta^{-}\partial_{\theta}^{3/0}+q^{2}%
\lambda\lambda_{+}^{-1}\theta^{3/0}\partial_{\theta}^{-}.\nonumber
\end{align}
The corresponding expressions for the conjugated calculus follow from the
above relations by applying the substitutions
\begin{equation}
\partial_{\theta}^{\mu}\rightarrow\hat{\partial}_{\theta}^{\bar{\mu}}%
,\quad\theta^{\mu}\rightarrow\bar{\theta}^{\bar{\mu}},\quad q\rightarrow
q^{-1}.
\end{equation}
As usual, the relations in (\ref{MinLeibRulAnf}) - (\ref{MinLeibRulEnd})
enable us to compute the action of fermionic derivatives on supernumbers:
\begin{align}
&  \partial_{\theta}^{+}\triangleright f(\theta^{-},\theta^{3/0},\theta
^{3},\theta^{+})\\
&  =-qf_{-}-qf_{-+}\theta^{+}-qf_{-3}\theta^{3}-q(f_{-,3/0}-\lambda\lambda
_{+}^{-1}f_{-3})\theta^{3/0}\nonumber\\
&  -qf_{-3+}\theta^{3}\theta^{+}-q(f_{-,3/0,+}-\lambda\lambda_{+}^{-1}%
f_{-3+})\theta^{3/0}\theta^{+}\nonumber\\
&  -qf_{-,3/0,3}\theta^{3/0}\theta^{3}-qf_{-,3/0,3+}\theta^{3/0}\theta
^{3}\theta^{+},\nonumber\\[0.16in]
&  \partial_{\theta}^{3/0}\triangleright f(\theta^{-},\theta^{3/0},\theta
^{3},\theta^{+})\\
&  =f_{3}+f_{3+}\theta^{+}-q^{2}f_{3/0,3}\theta^{3/0}-f_{-3}\theta
^{-}\nonumber\\
&  -q^{2}f_{3/0,3+}\theta^{3/0}\theta^{+}-f_{-3+}\theta^{-}\theta^{+}%
+q^{2}f_{-,3/0,3}\theta^{-}\theta^{3/0}\nonumber\\
&  +q^{2}f_{-,3/0,3+}\theta^{-}\theta^{3/0}\theta^{+},\nonumber\\[0.16in]
&  \partial_{\theta}^{0}\triangleright f(\theta^{-},\theta^{3/0},\theta
^{3},\theta^{+})\\
&  =f_{3/0}+f_{3/0,3}\theta^{3}+(f_{3/0,+}-f_{3+}\lambda\lambda_{+}%
^{-1})\theta^{+}\nonumber\\
&  -(q^{2}f_{-,3/0}+\lambda\lambda_{+}^{-1}f_{-3})\theta^{-}\nonumber\\
&  -q\lambda\lambda_{+}^{-1}(f_{-+}-qf_{3/0,3})\theta^{3/0}\nonumber\\
&  +f_{3/0,3+}\theta^{3}\theta^{+}+q\lambda f_{3/0,3+}\theta^{3/0}\theta
^{+}-q^{2}f_{-3,3/0,3}\theta^{-}\theta^{3}\nonumber\\
&  +q\lambda\lambda_{+}^{-1}f_{-3+}\theta^{3/0}\theta^{3}-q(qf_{-,3/0,+}%
-\lambda^{2}\lambda_{+}^{-1}f_{-3+})\theta^{-}\theta^{+}\nonumber\\
&  -q^{2}f_{-,3/0,3+}\theta^{-}\theta^{3}\theta^{+}-q^{2}\lambda\lambda
_{+}^{-1}f_{-,3/0,3+}\theta^{-}\theta^{3/0}\theta^{+},\nonumber\\[0.16in]
&  \partial_{\theta}^{-}\triangleright f(\theta^{-},\theta^{3/0},\theta
^{3},\theta^{+})\\
&  =-q^{-1}f_{+}+q^{-1}f_{3+}\theta^{3}+q(f_{+-}-\lambda f_{3/0,3})\theta
^{-}\nonumber\\
&  +(qf_{3/0,+}+q^{-1}\lambda\lambda_{+}^{-1}f_{3+})\theta^{3/0}\nonumber\\
&  -qf_{3/0,3+}\theta^{3/0}\theta^{3}-qf_{-3+}\theta^{-}\theta^{3}\nonumber\\
&  -q(q^{2}f_{-,3/0,+}+\lambda\lambda_{+}^{-1}f_{-3+})\theta^{-}\theta
^{3/0}\nonumber\\
&  -q\lambda f_{3/0,3+}\theta^{-}\theta^{+}+q^{3}f_{-,3/0,3+}\theta^{-}%
\theta^{3/0}\theta^{3}.\nonumber
\end{align}
The other types of representations of fermionic derivatives are completely
determined by transformation rules of the form
\begin{align}
&  \partial_{\theta}^{\mu}\triangleright f(\theta^{-},\theta^{3/0},\theta
^{3},\theta^{+})\\
\overset{{%
\genfrac{}{}{0pt}{}{+}{q}%
}{%
\genfrac{}{}{0pt}{}{\rightarrow}{\rightarrow}%
}{%
\genfrac{}{}{0pt}{}{-}{1/q}%
}}{\longleftrightarrow}  &  \hat{\partial}_{\theta}^{\bar{\mu}}\,\bar
{\triangleright}\,f(\bar{\theta}^{+},\bar{\theta}^{3/0},\bar{\theta}^{3}%
,\bar{\theta}^{-}),\nonumber\\[0.16in]
&  q^{2}f(\bar{\theta}^{-},\bar{\theta}^{3},\bar{\theta}^{3/0},\bar{\theta
}^{+})\,\bar{\triangleleft}\,\hat{\partial}_{\theta}^{\mu}\\
\overset{{%
\genfrac{}{}{0pt}{}{+}{q}%
}{%
\genfrac{}{}{0pt}{}{\rightarrow}{\rightarrow}%
}{%
\genfrac{}{}{0pt}{}{-}{1/q}%
}}{\longleftrightarrow}  &  q^{-2}f(\theta^{+},\theta^{3},\theta^{3/0}%
,\theta^{-})\triangleleft\partial_{\theta}^{\mu},\nonumber
\end{align}
and
\begin{align}
&  f(\bar{\theta}^{-},\bar{\theta}^{3},\bar{\theta}^{3/0},\bar{\theta}%
^{+})\,\bar{\triangleleft}\,\hat{\partial}_{\theta}^{\bar{\mu}}\\
\overset{+\leftrightarrow-}{\longleftrightarrow}  &  q^{-2}\partial_{\theta
}^{\mu}\triangleright f(\theta^{-},\theta^{3/0},\theta^{3},\theta
^{+}),\nonumber\\[0.16in]
&  f(\theta^{+},\theta^{3},\theta^{3/0},\theta^{-})\triangleleft
\partial_{\theta}^{\mu}\\
\overset{+\leftrightarrow-}{\longleftrightarrow}  &  q^{2}\hat{\partial
}_{\theta}^{\bar{\mu}}\,\bar{\triangleright}\,f(\bar{\theta}^{+},\bar{\theta
}^{3/0},\bar{\theta}^{3},\bar{\theta}^{-}),\nonumber
\end{align}
where
\begin{align}
\theta^{i_{1}}\cdots\theta^{i_{n}}  &  \overset{{%
\genfrac{}{}{0pt}{}{+}{q}%
}{%
\genfrac{}{}{0pt}{}{\rightarrow}{\rightarrow}%
}{%
\genfrac{}{}{0pt}{}{-}{1/q}%
}}{\longleftrightarrow}\theta^{\overline{i_{1}}}\cdots\theta^{\overline{i_{n}%
}},\\
f_{i_{1}\cdots i_{n}}  &  \overset{{%
\genfrac{}{}{0pt}{}{+}{q}%
}{%
\genfrac{}{}{0pt}{}{\rightarrow}{\rightarrow}%
}{%
\genfrac{}{}{0pt}{}{-}{1/q}%
}}{\longleftrightarrow}f_{\overline{i_{1}}\cdots\overline{i_{n}}},\nonumber\\
f^{\prime}  &  \overset{{%
\genfrac{}{}{0pt}{}{+}{q}%
}{%
\genfrac{}{}{0pt}{}{\rightarrow}{\rightarrow}%
}{%
\genfrac{}{}{0pt}{}{-}{1/q}%
}}{\longleftrightarrow}f^{\prime},\nonumber\\
q  &  \overset{{%
\genfrac{}{}{0pt}{}{+}{q}%
}{%
\genfrac{}{}{0pt}{}{\rightarrow}{\rightarrow}%
}{%
\genfrac{}{}{0pt}{}{-}{1/q}%
}}{\longleftrightarrow}q^{-1}.\nonumber
\end{align}

Finally, we come to the Hopf structure for the quantum spaces of the
q-deformed Lorentz-algebra. In general, we have
\begin{align}
\Delta(a^{3/0}) &  =a^{3/0}\otimes1+\Lambda(a)\tau^{1}\otimes a^{3/0}\\
&  -\,q^{1/2}\lambda_{+}^{1/2}\lambda\Lambda(a)(\tau^{3})^{-1/2}S^{1}\otimes
a^{+},\nonumber\\
\Delta(a^{+}) &  =a^{+}\otimes1+\Lambda^{1/2}(\tau^{3})^{-1/2}\sigma
^{2}\otimes a^{+}-q^{3/2}\lambda_{+}^{-1/2}\lambda\Lambda(a)T^{2}\otimes
a^{3/0},\nonumber\\
\Delta(a^{-}) &  =a^{-}\otimes1+\Lambda(a)(\tau^{3})^{1/2}\tau^{1}\otimes
a^{-}-q^{-1/2}\lambda_{+}^{1/2}\lambda\Lambda(a)S^{1}\otimes a^{0}\nonumber\\
&  -\,\lambda^{2}\Lambda(a)(\tau^{3})^{-1/2}T^{-}S^{1}\otimes a^{+}\nonumber\\
&  +\,q^{-1/2}\lambda_{+}^{-1/2}\lambda\Lambda(a)(\tau^{1}T^{-}-q^{-1}%
S^{1})\otimes a^{3/0},\nonumber\\
\Delta(a^{0}) &  =a^{0}\otimes1+\Lambda(a)\sigma^{2}\otimes a^{0}%
-q^{1/2}\lambda_{+}^{-1/2}\lambda\Lambda(a)T^{2}(\tau^{3})^{1/2}\otimes
a^{-}\nonumber\\
&  +\,q^{1/2}\lambda_{+}^{-1/2}\lambda\Lambda(a)(\tau^{3})^{-1/2}(T^{-}%
\sigma^{2}+qS^{1})\otimes a^{+}\nonumber\\
&  -\,\lambda_{+}^{-1}\Lambda(a)(\lambda^{2}T^{-}T^{2}+q(\tau^{1}-\sigma
^{2}))\otimes a^{3/0},\nonumber\\[0.16in]
S(a^{3/0}) &  =-\Lambda^{-1}(a)\sigma^{2}a^{3/0}-q^{-3/2}\lambda_{+}%
^{1/2}\lambda\Lambda^{-1}(a)S^{1}a^{+},\\
S(a^{+}) &  =-\Lambda^{-1}(a)\tau^{1}(\tau^{3})^{1/2}a^{+}-q^{3/2}\lambda
_{+}^{-1/2}\lambda\Lambda^{-1}(a)T^{2}(\tau^{3})^{1/2}a^{3/0},\nonumber\\
S(a^{-}) &  =-\Lambda^{-1}(a)\sigma^{2}(\tau^{3})^{-1/2}a^{-}-q^{-1/2}%
\lambda_{+}^{1/2}\lambda\Lambda^{-1}(a)(\tau^{3})^{-1/2}S^{1}a^{0}\nonumber\\
&  +\,q^{-2}\lambda^{2}\Lambda^{-1}(a)(\tau^{3})^{-1/2}S^{1}T^{-}%
a^{+}\nonumber\\
&  +\,q^{-5/2}\lambda_{+}^{-1/2}\lambda\Lambda^{-1}(a)(\tau^{3})^{-1/2}%
(\sigma^{2}T^{-}-q^{3}S^{1})a^{3/0},\nonumber\\
S(a^{0}) &  =-\Lambda^{-1}(a)\tau^{1}a^{0}-q^{5/2}\lambda_{+}^{-1/2}%
\lambda\Lambda^{-1}(a)T^{2}a^{-}\nonumber\\
&  +\,q^{-3/2}\lambda_{+}^{-1/2}\lambda\Lambda^{-1}(a)(\tau^{1}T^{-}%
+qS^{1})a^{+}\nonumber\\
&  +\,\lambda_{+}^{-1}\Lambda^{-1}(a)(q(\sigma^{2}-\tau^{1})+\lambda^{2}%
T^{2}T^{-})a^{3/0},\nonumber\\[0.16in]
\varepsilon(a^{3/0}) &  =\varepsilon(a^{+})=\varepsilon(a^{-})=\varepsilon
(a^{0})=0,
\end{align}
and likewise for the conjugated Hopf structure,
\begin{align}
\bar{\Delta}(a^{3/0}) &  =a^{3/0}\otimes1+\Lambda^{-1}(a)(\tau^{3}%
)^{1/2}\sigma^{2}\otimes a^{3/0}\\
&  -q^{3/2}\lambda_{+}^{1/2}\lambda\Lambda^{-1}(a)T^{2}\otimes a^{-}%
,\nonumber\\
\bar{\Delta}(a^{-}) &  =a^{-}\otimes1+\Lambda^{-1}(a)\tau^{1}\otimes
a^{-}-q^{1/2}\lambda_{+}^{-1/2}\lambda\Lambda^{-1}(a)(\tau^{3})^{-1/2}%
S^{1}\otimes a^{3/0},\nonumber\\
\bar{\Delta}(a^{+}) &  =a^{+}\otimes1+\Lambda^{-1}(a)\sigma^{2}\otimes
a^{+}-q^{1/2}\lambda_{+}^{1/2}\lambda\Lambda^{-1}(a)T^{2}(\tau^{3}%
)^{1/2}\otimes a^{0}\nonumber\\
&  -q^{1/2}\lambda_{+}^{-1/2}\lambda\Lambda^{-1}(a)(\tau^{3})^{-1/2}%
(T^{+}\sigma^{2}+q\tau^{3}T^{2})\otimes a^{3/0}\nonumber\\
&  +q^{2}\lambda^{2}\Lambda^{-1}(a)T^{2}T^{+}\otimes a^{-},\nonumber\\
\bar{\Delta}(a^{0}) &  =a^{0}\otimes1+\Lambda^{-1}(a)(\tau^{3})^{-1/2}\tau
^{1}\otimes a^{0}-q^{-1/2}\lambda_{+}^{-1/2}\lambda\Lambda^{-1}(a)S^{1}\otimes
a^{+}\nonumber\\
&  -q^{1/2}\lambda_{+}^{-1/2}\lambda\Lambda^{-1}(a)(qT^{+}\tau^{1}%
-T^{2})\otimes a^{-}\nonumber\\
&  +\lambda_{+}^{-1}\Lambda^{-1}(a)(\tau^{3})^{-1/2}(\lambda^{2}T^{+}%
S^{1}+q^{-1}(\tau^{3}\tau^{1}-\sigma^{2}))\otimes a^{3/0},\nonumber\\[0.16in]
\bar{S}(a^{3/0}) &  =-\Lambda(a)\tau^{1}(\tau^{3})^{1/2}a^{3/0}-q^{3/2}%
\lambda_{+}^{1/2}\lambda\Lambda(a)T^{2}(\tau^{3})^{1/2}a^{-},\\
\bar{S}(a^{-}) &  =-\Lambda(a)\sigma^{2}a^{-}-q^{-3/2}\lambda_{+}%
^{-1/2}\lambda\Lambda(a)S^{1}a^{3/0},\nonumber\\
\bar{S}(a^{+}) &  =-\Lambda(a)\tau^{1}a^{+}-q^{5/2}\lambda_{+}^{1/2}%
\lambda\Lambda(a)T^{2}a^{0}\nonumber\\
&  -q^{3/2}\lambda_{+}^{-1/2}\lambda\Lambda(a)(q\tau^{1}T^{+}+T^{2}%
)a^{3/0}\nonumber\\
&  -q^{4}\lambda^{2}\Lambda(a)T^{2}T^{+}a^{-},\nonumber\\
\bar{S}(a^{0}) &  =-\Lambda(a)(\tau^{3})^{-1/2}\sigma^{2}a^{0}-q^{-1/2}%
\lambda_{+}^{-1/2}\lambda\Lambda(a)(\tau^{3})^{-1/2}S^{1}a^{+}\nonumber\\
&  -q^{3/2}\lambda_{+}^{-1/2}\lambda\Lambda(a)(\tau^{3})^{-1/2}(\sigma
^{2}T^{+}-q\tau^{3}T^{2})a^{-}\nonumber\\
&  -\lambda_{+}^{-1}\Lambda(a)(\tau^{3})^{-1/2}(\lambda^{2}T^{+}S^{1}%
+q(\sigma^{2}-\tau^{3}\tau^{1}))a^{3/0},\nonumber\\[0.16in]
\bar{\varepsilon}(a^{3/0}) &  =\bar{\varepsilon}(a^{+})=\bar{\varepsilon
}(a^{-})=\bar{\varepsilon}(a^{0})=0,
\end{align}
with
\begin{equation}
a\in\{\partial_{x},\partial_{\theta},X,\theta,\xi,\eta\}.
\end{equation}
The scaling operators have to take the form
\begin{equation}
\Lambda(\partial_{x}^{i})=\Lambda^{1/2},\quad\Lambda(X^{i})=\Lambda
^{-1/2},\quad\Lambda(\eta^{i})=\Lambda^{-1},
\end{equation}
or
\begin{equation}
\Lambda(\partial_{\theta}^{i})=\tilde{\Lambda}^{-1},\quad\Lambda(\theta
^{i})=\tilde{\Lambda},\quad\Lambda(\xi^{i})=\tilde{\Lambda},
\end{equation}
if the operators $\Lambda$ and $\tilde{\Lambda}$ are subject to the relations%
\begin{align}
\Lambda X^{\mu} &  =q^{-2}X^{\mu}\Lambda, & \tilde{\Lambda}X^{\mu} &
=q^{-1}X^{\mu}\tilde{\Lambda},\\
\Lambda\partial_{x}^{\mu} &  =q^{2}\partial_{x}^{\mu}\Lambda, & \tilde
{\Lambda}\partial_{x}^{\mu} &  =q\partial_{x}^{\mu}\tilde{\Lambda},\nonumber\\
\Lambda\xi^{\mu} &  =q^{-2}\xi^{\mu}\Lambda, & \tilde{\Lambda}\xi^{\mu} &
=-q^{-1}\xi^{\mu}\tilde{\Lambda},\nonumber\\
\Lambda\eta^{\mu} &  =q^{-4}\eta^{\mu}\Lambda & \tilde{\Lambda}\eta^{\mu} &
=q^{2}\eta^{\mu}\tilde{\Lambda},\nonumber\\
\Lambda\theta^{\mu} &  =q^{-2}\theta^{\mu}\Lambda, & \tilde{\Lambda}%
\theta^{\mu} &  =-q^{-1}\theta^{\mu}\tilde{\Lambda},\nonumber\\
\Lambda\partial_{\theta}^{\mu} &  =q^{2}\partial_{\theta}^{\mu}\Lambda, &
\tilde{\Lambda}\partial_{\theta}^{\mu} &  =-q\partial_{\theta}^{\mu}%
\tilde{\Lambda}.\nonumber
\end{align}

\noindent\textbf{Acknowledgement}\newline First of all we want to express our
gratitude to Julius Wess for his efforts, suggestions and discussions. Also we
would like to thank Fabian Bachmaier for useful discussions and his steady support.

\appendix

\section{Representations of supernumbers\label{RepSup}}

In this article we deal with supernumbers of different normal orderings. Thus,
it can be useful to have formulae at hand that allow to switch between the
different orderings. For this purpose we wish to list the following identities:

\begin{enumerate}
\item (two-dimensional Euclidean space)\textbf{ }
\begin{align}
&  f^{\prime}+f_{1}\theta^{1}+f_{2}\theta^{2}+f_{12}\theta^{1}\theta^{2}\\
&  =\tilde{f}^{\prime}+\tilde{f}_{1}\theta^{1}+\tilde{f}_{2}\theta^{2}%
+\tilde{f}_{21}\theta^{2}\theta^{1},\nonumber
\end{align}
where
\begin{equation}
\tilde{f}^{\prime}=f^{\prime},\quad\tilde{f}_{1}=f_{1},\quad\tilde{f}%
_{2}=f_{2},\quad\tilde{f}_{21}=-q^{-1}f_{12}.
\end{equation}

\item (three-dimensional Euclidean space)\textbf{ }
\begin{align}
&  f^{\prime}+f_{+}\theta^{+}+f_{3}\theta^{3}+f_{-}\theta^{-}\\
&  +f_{+3}\theta^{+}\theta^{3}+f_{+-}\theta^{+}\theta^{-}+f_{3-}\theta
^{3}\theta^{-}\nonumber\\
&  +f_{+3-}\theta^{+}\theta^{3}\theta^{-}\nonumber\\
&  =\tilde{f}^{\prime}+\tilde{f}_{+}\theta^{+}+\tilde{f}_{3}\theta^{3}%
+\tilde{f}_{-}\theta^{-}\nonumber\\
&  +\tilde{f}_{3+}\theta^{3}\theta^{+}+\tilde{f}_{-+}\theta^{-}\theta
^{+}+\tilde{f}_{-3}\theta^{-}\theta^{3}\nonumber\\
&  +\tilde{f}_{-3+}\theta^{-}\theta^{3}\theta^{+},\nonumber
\end{align}
where
\begin{gather}
\tilde{f}^{\prime}=f^{\prime},\quad\tilde{f}_{A}=f_{A},\quad A\in\{+,3,-\}\\
\tilde{f}_{-+}=-f_{+-},\quad\tilde{f}_{3+}=-q^{2}f_{+3},\quad\tilde{f}%
_{-3}=-q^{2}f_{3-},\nonumber\\
\tilde{f}_{-3+}=-q^{4}f_{+3-}.\nonumber
\end{gather}

\item (four-dimensional Euclidean space)
\begin{align}
&  f^{\prime}+\sum_{i=1}^{4}f_{i}\theta^{i}+\sum_{1\leq i_{1}<i_{2}\leq
4}f_{i_{1}i_{2}}\theta^{i_{1}}\theta^{i_{2}}\\
&  +\sum_{1\leq i_{1}<i_{2}<i_{3}\leq4}f_{i_{1}i_{2}i_{3}}\theta^{i_{1}}%
\theta^{i_{2}}\theta^{i_{3}}+f_{1234}\theta^{1}\theta^{2}\theta^{3}\theta
^{4}\nonumber\\
&  =\tilde{f}^{\prime}+\sum_{i=1}^{4}\tilde{f}_{i}\theta^{i}+\sum_{1\leq
i_{2}<i_{1}\leq4}\tilde{f}_{i_{1}i_{2}}\theta^{i_{1}}\theta^{i_{2}}\nonumber\\
&  +\sum_{1\leq i_{3}<i_{2}<i_{1}\leq4}\tilde{f}_{i_{1}i_{2}i_{3}}%
\theta^{i_{1}}\theta^{i_{2}}\theta^{i_{3}}+\tilde{f}_{4321}\theta^{4}%
\theta^{3}\theta^{2}\theta^{1},\nonumber
\end{align}
where
\begin{gather}
\tilde{f}^{\prime}=f^{\prime},\quad\tilde{f}_{i}=f_{i},\quad i=1,\ldots,4,\\
\tilde{f}_{21}=-q^{-1}f_{12},\quad\tilde{f}_{31}=-q^{-1}f_{13},\nonumber\\
\tilde{f}_{41}=-f_{14}-\lambda f_{23},\quad\tilde{f}_{32}=-f_{23},\nonumber\\
\tilde{f}_{42}=-q^{-1}f_{24},\quad\tilde{f}_{43}=-q^{-1}f_{34},\nonumber\\
\tilde{f}_{321}=-q^{-2}f_{123},\quad\tilde{f}_{421}=-q^{-2}f_{124},\nonumber\\
\tilde{f}_{431}=-q^{-2}f_{134},\quad\tilde{f}_{432}=-q^{-2}f_{234},\nonumber\\
\tilde{f}_{4321}=q^{-4}f_{1234}.\nonumber
\end{gather}

\end{enumerate}

(Minkowski space)
\begin{align}
&  f^{\prime}+f_{+}\theta^{+}+f_{3/0}\theta^{3/0}+f_{3}\theta^{3}+f_{-}%
\theta^{-}\\
&  +f_{+,3/0}\theta^{+}\theta^{3/0}+f_{+3}\theta^{+}\theta^{3}+f_{+-}%
\theta^{+}\theta^{-}\nonumber\\
&  +f_{3/0,3}\theta^{3/0}\theta^{3}+f_{3-}\theta^{3}\theta^{-}+f_{3/0,3}%
\theta^{3/0}\theta^{3}+f_{3/0,-}\theta^{3/0}\theta^{-}\nonumber\\
&  +f_{+,3/0,3}\theta^{+}\theta^{3/0}\theta^{3}+f_{+3-}\theta^{+}\theta
^{3}\theta^{-}+f_{+,3/0,-}\theta^{+}\theta^{3/0}\theta^{-}\nonumber\\
&  +f_{3/0,3-}\theta^{3/0}\theta^{3}\theta^{-}+f_{+,3/0,3-}\theta^{+}%
\theta^{3/0}\theta^{3}\theta^{-}\nonumber\\
&  =\tilde{f}^{\prime}+\tilde{f}_{+}\theta^{+}+\tilde{f}_{3/0}\theta
^{3/0}+\tilde{f}_{3}\theta^{3}+\tilde{f}_{-}\theta^{-}\nonumber\\
&  +\tilde{f}_{3/0,+}\theta^{3/0}\theta^{+}+\tilde{f}_{3+}\theta^{3}\theta
^{+}+\tilde{f}_{-+}\theta^{-}\theta^{+}\nonumber\\
&  +\tilde{f}_{3,3/0}\theta^{3}\theta^{3/0}+\tilde{f}_{-3}\theta^{-}\theta
^{3}+\tilde{f}_{-,3/0}\theta^{-}\theta^{3/0}\nonumber\\
&  +\tilde{f}_{3,3/0,+}\theta^{3}\theta^{3/0}\theta^{+}+\tilde{f}_{-3+}%
\theta^{-}\theta^{3}\theta^{+}+\tilde{f}_{-,3/0,+}\theta^{-}\theta^{3/0}%
\theta^{+}\nonumber\\
&  +\tilde{f}_{-3,3/0}\theta^{-}\theta^{3}\theta^{3/0}+\tilde{f}%
_{-3,3/0,+}\theta^{-}\theta^{3}\theta^{3/0}\theta^{+},\nonumber
\end{align}
where
\begin{gather}
\tilde{f}^{\prime}=f^{\prime},\quad\tilde{f}_{\mu}=f_{\mu},\quad\mu
\in\{+,3/0,3,-\},\\
\tilde{f}_{+3}=-q^{-2}f_{3+},\quad\tilde{f}_{+,3/0}=-f_{3/0,+},\nonumber\\
\tilde{f}_{+-}=-f_{-+}+\lambda f_{3/0,3},\quad\tilde{f}_{3,3/0}=-f_{3/0,3}%
,\nonumber\\
\tilde{f}_{3-}=-q^{-2}f_{-3},\quad\tilde{f}_{3/0,-}=-f_{-,3/0}\nonumber\\
\tilde{f}_{+3,3/0}=-q^{2}f_{3/0,3+},\quad\tilde{f}_{+3-}=-q^{-4}%
f_{-3+},\nonumber\\
\tilde{f}_{+,3/0,-}=-f_{-,3/0,+},\quad\tilde{f}_{3,3/0,-}=-q^{-2}%
f_{-,3/0,3},\nonumber\\
\tilde{f}_{+3,3/0,-}=q^{-4}f_{-,3/0,3+}.\nonumber
\end{gather}


\begin{thebibliography}{99}                                                                                               %
\bibitem {Heis38}W. Heisenberg, \textit{\"{U}ber die in der Theorie der
Elementarteilchen auftretende universelle L\"{a}nge, }Ann. Phys. \textbf{32},
20 (1938).

\bibitem {Sny47}H.S. Snyder, \textit{Quantized space-time}, Phys. Rev.
\textbf{71}, 38 (1947)

\bibitem {Ku83}P.P. Kulish, N.Y. Reshetikhin, \textit{Quantum linear problem
for the Sine-Gordon equation and higher representations, }J. Sov. Math.
\textbf{23}, 2345 (1983)

\bibitem {Drin86}V.G. Drinfeld, \textit{Quantum groups}, in
\textit{Proceedings of the International Congress of Mathematicians, 1986},
edited by A. M. Gleason (Amer. Math. Soc., 1986), p. 798

\bibitem {Dri85}V.G. Drinfeld, \textit{Hopf algebras and the quantum
Yang-Baxter equation, }Sov. Math. Dokl. \textbf{32}, 254 (1985)

\bibitem {Jim85}M. Jimbo, \textit{A q-analogue of U(g) and the Yang-Baxter
equation,} Lett. Math. Phys. \textbf{10}, 63 (1985)

\bibitem {Wor87}S.L. Woronowicz, \textit{Compact matrix pseudo groups,
}Commun. Math. Phys. \textbf{111}, 613 (1987)

\bibitem {Man88}Y.J. Manin, \textit{Quantum groups and Non-Commutative
Geometry }(Centre de Recherche Mathematiques, Montreal 1988)

\bibitem {RFT90}N.Yu. Reshetikhin, L.A. Takhtadzhyan, L.D. Faddeev,
\textit{Quantization of Lie Groups and Lie Algebras}, Leningrad Math. J.
\textbf{1}, 193 (1990)

\bibitem {CSSW90}U. Carow-Watamura, M. Schlieker, M. Scholl, S. Watamura,
\textit{Tensor Representations of the Quantum Group }$SL_{q}(2)$ \textit{and
Quantum Minkowski Space, }Z. Phys. C \textbf{48}, 159 (1990)

\bibitem {Pod90}P. Podles, S.L. Woronowicz, \textit{Quantum Deformation of
Lorentz Group, }Commun. Math. Phys. \textbf{130}, 381 (1990)

\bibitem {SWZ91}W.B. Schmidke, J. Wess, B. Zumino, \textit{A q-deformed
Lorentz Algebra, }Z. Phys. C \textbf{52}, 471 (1991)

\bibitem {Maj91}S. Majid, \textit{Examples of braided groups and braided
matrices,} J. Math. Phys. \textbf{32}, 3246 (1991)

\bibitem {LWW97}A. Lorek, W. Weich, J. Wess, \textit{Non-Commutative Euclidean
and Minkowski Structures, }Z. Phys. C \textbf{76}, 375\ (1997), [q-alg/9702025]

\bibitem {Fich97}M. Fichtm\"{u}ller, A. Lorek, J. Wess, \textit{q-deformed
Phase Space and its Lattice Structure},\textit{ }Z. Phys. C \textbf{71}, 533 (1996)

\bibitem {CW98}B.L. Cerchiai, J. Wess, \textit{q-Deformed Minkowski Space
based on a q-Lorentz Algebra, }Eur. Phys. J. C \textbf{5}, 553 (1998), [math.QA/9801104]

\bibitem {MajReg}S. Majid, \textit{On the q-regularisation, }Int. J. Mod.
Phys. A \textbf{5}, 4689 (1990)

\bibitem {GKP96}H. Grosse, C. Klim\v{c}ik, P. Pre\v{s}najder, \textit{Towards
finite quantum field theory in non-commutative geometry, }Int. J. Theor. Phys.
\textbf{35}, 231 (1996), [hep-th/9505175].

\bibitem {Oec99}R. Oeckl, \textit{Braided Quantum Field Theory, }Commun. Math.
Phys. \textbf{217},\textbf{ }451 (2001)

\bibitem {Blo03}C. Blohmann, \textit{Free q-deformed relativistic wave
equations by representation theory}, Eur. Phys. J. C \textbf{30}, 435 (2003), [hep-th/0111172]

\bibitem {WW01}H. Wachter, M. Wohlgenannt, $\ast$\textit{-Products on quantum
spaces, }Eur. Phys. J. C \textbf{23}, 761 (2002),\textit{\ }[hep-th/0103120]

\bibitem {BW01}C. Bauer, H. Wachter, \textit{Operator representations on
quantum spaces, }Eur. Phys. J. C \textbf{31}, 261 (2003), [math-ph/0201023]

\bibitem {Wac02}H. Wachter, \textit{q-Integration on quantum spaces, }Eur.
Phys. J. C \textbf{32}, 281 (2004), [hep-th/0206083]

\bibitem {Wac04}H. Wachter, \textit{q-Exponentials on quantum spaces}, Eur.
Phys. J. C \textbf{37}, 379 (2004), [hep-th/0401113]

\bibitem {WacTr}H. Wachter, \textit{q-Translations on quantum spaces,
}preprint, [hep-th/0410205]

\bibitem {WacBr}H. Wachter, \textit{Braided products for quantum spaces,
}preprint, [math-ph/0509018]

\bibitem {Maj95}S. Majid, \textit{Foundations of Quantum Group Theory}
(University Press, Cambridge 1995)

\bibitem {ChDe96}M. Chaichian, A.P. Demichev, \textit{Introduction to Quantum
Groups} (World Scientific, Singapore, 1996)

\bibitem {Maj91Kat}S. Majid, \textit{Representations, duals and quantum
doubles of monoidal categories, }Suppl. Rend. Circ. Mat. Palermo, Ser. II,
\textbf{26}, 197 (1991)

\bibitem {Maj94Kat}S. Majid, \textit{Algebras and Hopf Algebras in Braided
Categories, }Lec. Notes Pure Appl. Math. \textbf{158}, 55 (1994)

\bibitem {MaL74}S. Mac Lane, \textit{Categories for the Working Mathematician}
(Springer, 1974)

\bibitem {KS97}A. Klimyk, K. Schm\"{u}dgen, \textit{Quantum Groups and their
Representations} (Springer, Berlin 1997)

\bibitem {LSW94}A. Lorek, W.B. Schmidke, J. Wess, $SU_{q}(2)$
\textit{Covariant }$\hat{R}$\textit{-Matrices for Reducible Representations,}
Lett. Math. Phys. \textbf{31},\textbf{ }279 (1994)

\bibitem {Wess00}J. Wess, \textit{q-deformed Heisenberg Algebras}, in
\textit{Proceedings of the 38. Internationale Universit\"{a}tswochen f\"{u}r
Kern- und Teilchenphysik, Lect. Notes in Phys. no. 543, Schladming, 2000},
edited by H. Gausterer, H. Grosse, L. Pittner (Springer, 2000), [math-phy/9910013]

\bibitem {WZ91}J. Wess, B. Zumino,\textit{\ Covariant differential calculus on
the quantum hyperplane}, Nucl. Phys. B Suppl. \textbf{18}, 302 (1991)

\bibitem {CSW91}U. Carow-Watamura, M. Schlieker, S. Watamura, $SO_{q}%
(N)$\textit{-covariant differential calculus on quantum space and deformation
of Schr\"{o}dinger equation,} Z. Phys. C \textbf{49}, 439 (1991)

\bibitem {Song92}X. C. Song, \textit{Covariant differential calculus on
quantum minkowski space and q-analog of Dirac equation, }Z. Phys. C
\textbf{55}, 417 (1992)

\bibitem {Tan3}T. Tanisaki, \textit{Killing Forms, Harish-Chandra
homomorphisms and universal R-matrices for quantum algebras, }in
\textit{Infinite Algebras}, edited by A. Tsuchiya, T. Eguchi, M. Jimbo (World
Scientific, Singapore 1992)

\bibitem {Maj-Pr}S. Majid, \textit{Braided momentum in the
q-Poincar\'{e}-group,} J. Math. Phys. \textbf{34}, 2054 (1993), [hep-th/9210141]

\bibitem {Maj94star}S. Majid, $\mathit{\ast}$\textit{-structures on braided
spaces,} J. Math. Phys. \textbf{36}, 4436 (1995)

\bibitem {Maj95star}S. Majid, \textit{Quasi-}$\mathit{\ast}$\textit{-structure
on q-Poincar\'{e} algebras,} J. Geom. Phys. \textbf{22}, 14 (1997)

\bibitem {SS90}M. Schlieker, W. Scholl, \textit{Spinor calculus for quantum
groups, }Z. Phys. C \textbf{52}, 471 (1991)

\bibitem {Oca96}H. Ocampo, $SO_{q}(4)$ \textit{quantum mechanics}, Z. Phys. C
\textbf{70}, 525 (1996)

\bibitem {OSWZ92}O. Ogievetsky, W.B. Schmidke, J. Wess, B. Zumino,
\textit{q-Deformed Poincar\'{e} Algebra,} Commun. Math. Phys. \textbf{150},
495 (1992)

\bibitem {Lu92}J. Lukierski, A. Nowicki, H. Ruegg, \textit{New Quantum
Poincare Algebra and }$\kappa$\textit{-deformed Field Theory, }Phys. Lett.
B\textbf{ 293}, 344 (1992)

\bibitem {Cas93}L. Castellani, \textit{Differential Calculus on }$ISO_{q}%
(N)$\textit{, Quantum Poincar\'{e} Algebra and q-Gravity,} preprint, [hep-th/9312179]

\bibitem {Dov94}V.K. Dobrev, \textit{New q-Minkowski space-time and q-Maxwell
equations hierarchy from q-conformal invariance,} Phys. Lett. B \textbf{341},
133 (1994)

\bibitem {ChDe95}M. Chaichian, A.P. Demichev, \textit{Quantum Poincar\'{e}
group without dilatation and twisted classical algebra}, J. Math. Phys.
\textbf{36}, 398 (1995)

\bibitem {ChKu04}M. Chaichian, P.P. Kulish, K. Nishijima, A. Tureanu,
\textit{On a Lorentz-Invariant Interpretation of Noncommutative Space-Time and
its Implications on Noncommutative QFT, }Phys. Lett. B \textbf{604}, 98
(2004), [hep-th/0408062]

\bibitem {Koch04}F. Koch, E. Tsouchnika, \textit{Construction of }$\theta
$-\textit{Poincar\'{e} Algebras and their invariants on }$\mathcal{M}_{\theta
}$, Nucl. Phys. B \textbf{717}, 387 (2005),\textbf{ }[hep-th/0409012]
\end{thebibliography}
\end{document}